\documentclass[11pt]{article}
\usepackage{amsmath}
\usepackage{amssymb}
\usepackage{color}
\usepackage{graphicx}
\newtheorem{defn1}{Definition}[section]

\newtheorem{prop1}[defn1]{Proposition}
\newtheorem{thm1}[defn1]{Theorem}

\newtheorem{rmk1}[defn1]{Remark}

\newcommand{\eps}{\varepsilon}
\newcommand{\reals}{\mathbb{R}}

\newcommand{\mapping}{\longrightarrow}

\renewcommand{\d}{\mathrm{d}}
\newcommand{\grad}{\nabla}

\newcommand{\Tang}[1]{\mathrm{T}{#1}}
\newcommand{\Diff}[1]{\frac{\partial}{\partial{#1}}}
\newcommand{\diff}[2]{\frac{\partial{#1}}{\partial{#2}}}

\newcommand{\inner}[2]{\left\langle #1,#2\right\rangle}

\newcommand{\X}{\mathcal{X}}
\newcommand{\id}{\,{\mathrm{l}\!\!\!{1}}}
\newcommand{\Form}[2]{{\Omega^{#2}}(#1)}
\newcommand{\sphere}[1]{{\mathrm{S}^{#1}}}
\newcommand{\bu}{{\mbox{\boldmath$u$}}}
\newcommand{\bx}{{\mbox{\boldmath$x$}}}
\newcommand{\bm}{{\mbox{\boldmath$m$}}}
\newcommand{\dr}{{\d{r}}}

\newcommand{\SO}[1]{{\mathrm {SO}}(#1)}

\newcommand{\SL}[2]{{\mathrm {SL}}(#1,#2)}

\newcommand{\hyperbolic}{\mathcal{H}}

\newcommand{\proj}{\mathbb{P}}

\newcommand{\comment}[1]{\par
\framebox{\begin{minipage}[c]{.95 \textwidth} \rm #1
\end{minipage}}\vspace{1 mm}\par}
\newcommand{\rem}[1]{}

\begin{document}
\title{Reduced Singular Solutions of EPDiff Equations\\ on Manifolds
with Symmetry}
\author{Darryl D. Holm$^{1,2}$, Jonathan Munn$^{2}$ and Samuel N. Stechmann$^{1,3}$\\ \\
$^{1}$ Computational and Computer Science Division\\
 Los Alamos National Laboratory,\\
 MS D413 Los Alamos\\
 NM 87545 USA\\
\small{email: dholm@lanl.gov}\\ \\
 $^{2}$ Mathematics Department\\
 Imperial College\\
SW7 2AZ London UK\\
   \small{email: d.holm@imperial.ac.uk}\\
   \small{email: j.munn@imperial.ac.uk}\\
\\
$^3$Mathematics Department, Courant Institute\\
New York University\\
251 Mercer St.\\
New York, NY 10012\\
  \small{email: stechman@cims.nyu.edu}\\
\date{
\comment{\begin{center}
I'll put a girdle round about the earth in forty minutes.
\\ -- Puck, in {\it A Midsummer Night's Dream}, Act II, scene i.
\end{center}}
August 10, 2004}}
\maketitle
\begin{abstract}
The EPDiff equation governs geodesic flow on the diffeomorphisms with
respect to a chosen metric, which is typically a Sobolev norm on the
tangent space of vector fields. EPDiff admits a remarkable ansatz for its
singular solutions, called ``diffeons,'' whose momenta are supported on
embedded subspaces of the ambient space. Diffeons are true solitons for some
choices of the norm. The diffeon solution ansatz is a momentum map.
Consequently. the diffeons evolve according to canonical Hamiltonian equations.
We examine diffeon solutions on Einstein spaces that are ``mostly" symmetric,
i.e., whose quotient by a subgroup of the isometry group is 1-dimensional. An
example is the two-sphere, whose isometry group $\SO{3}$ contains $S^1$. In
this situation, the singular diffeons (called ``Puckons'') are
supported on latitudes (``girdles'') of the sphere. For this $S^1$ symmetry of
the two-sphere, the canonical Hamiltonian dynamics for Puckons reduces from
integral partial differential equations to a dynamical system of ordinary
differential equations for their colatitudes.  Explicit examples are computed
numerically for the motion and interaction of the Puckons on the sphere with
respect to the $H^1$ norm. We analyse this case and several other 2-dimensional
examples. From consideration of these 2-dimensional spaces, we outline the
theory for reduction of diffeons on a general manifold possessing a metric
equivalent to the warped product of the line with the bi-invariant metric of a
Lie group. 
\rem{Finally,
we consider the reduction to finite dimensional diffeon solutions for 
two-dimensional Teichm\"uller spaces with respect to the Weil-Petersson
norm.} 
\end{abstract}

\tableofcontents
\section{Introduction}
\subsection{Motivation and Problem Statement}

When the Schumacher-Levi comet broke into a string of fragments
that collided with Jupiter in July 1994, each of the impacts created a
circularly symmetric expanding ``ripple'' spreading out into the gaseous sphere
of Jupiter. If the collision had occurred instead with an Earth-like planet
which was entirely covered by a shallow layer of water, it would have
produced a sequence of strongly nonlinear circular traveling-wave ripples
spreading out from a point source on the spherical surface, propagating to the
antipodal point, then reflecting back in the opposite direction and colliding
head-on with the trailing waves. \par

Neglecting planetary rotation and assuming the energy was
mainly kinetic in this situation would produce an approximate
description of these strongly nonlinear shallow water waves as solitary
waves governed by the Camassa-Holm (CH) equation \cite{CaHo1993}, written
on the sphere. The momenta for these solitary waves on the surface of the
sphere would be concentrated on a series of coaxial circles which could
each be regarded as a latitude encircling the axis running from the point
of impact to the antipodal point. These concentrations of momentum on
latitudes are singular solutions of the EPDiff equation, which is a
geometrical extension of the CH equation for shallow water waves to higher
dimensions.
\par

Previously, the CH equation and its extension EPDiff have been solved
analytically for the interactions of such waves on the real line in one
dimension \cite{CaHo1993}, as radially-symmetric rotating concentric circles on
the two-dimensional plane \cite{HoPuSt2004} and numerically in two dimensions
and three dimensions, for a variety of initial value problems
\cite{HoSt2004}. In cases with one-dimensional linear, or radial symmetry,
these dynamics reduce to ordinary differential equations (ODEs). Here we
consider the corresponding reduction for singular wave motion on the surface of
the sphere for EPDiff with respect to the $H^1$ norm and we discuss its
generalisations for other surfaces of constant curvature.
\par

In each case, we derive the symmetry which reduces the dynamics of the singular
solutions of the EPDiff equation to a system of ODEs. We also provide explicit
numerical results for the interactions of these singular EPDiff solutions in
the case of Puckon motion as concentric latitudes moving on the sphere.

\subsection{Review and Generalisation of the Camassa-Holm Equations to
EPDiff}\label{Intro}
The dispersionless Camassa-Holm (CH) equations \cite{CaHo1993} for
one-dimensional shallow water waves arise as stationary points of the
kinetic energy functional given by the Sobolev (2,1)-norm
\begin{eqnarray*}
\ell &:&W^{2,1}(\reals)\mapping\reals\,,\\
\ell[u]&=&\int_\reals \left(u(x)^2+u_x(x)^2\right)\d x\,,
\end{eqnarray*}
subject to velocity variations in the Euler-Poincar\'e form \cite{HoMaRa1998}
\[
\delta u=\dot{\zeta}+[\zeta,u]\,,
\]
where $u,\zeta\in\mathfrak{g}$, and $\mathfrak{g}$ consists of the vector
fields on the real line, with Lie bracket $[\cdot,\cdot]$ given by
\[
[v,w]=vw_x-wv_x=\mathrm{ad}_vw\,.
\]
The dispersionless CH equations themselves are thus given by the following
system for velocity $u$ and its dual momentum $m$,
\begin{eqnarray*}
\dot{m}&=&-\mathrm{ad}_u^*m=-(m\partial_x+\partial_x m)u
\,,\\
\hbox{where}\quad
m&=&\frac{\delta \ell}{\delta u}
=u-u_{xx}=u+\Delta u
\,,\\
\Longrightarrow\quad u&=&G*m=\int_\reals G(x-y)m(y)dy\,,
\end{eqnarray*}
and $G(x)=\frac{1}{2}e^{-|x|}$ is the Green's function for the Helmholtz
operator $1-\partial_x^2$ on the real line. We remark that here we are
using the convention that the Laplacian in 1D is given by
\[
\Delta u =-u_{xx}.
\]
The sign comes from regarding the Laplacian as
$\partial_x^*\partial_x$, where $\partial_x^*$ is the $L^2$ adjoint of the
derivative operator $\partial_x$.
\par 

To put the dispersionless CH equations onto a general Riemannian manifold
$(M,\inner{\cdot}{\cdot})$ with dim$M=n$ and Levi-Civita connection $\grad$, we
work with the functional on the space of weakly differentiable
square-integrable vector fields
\begin{eqnarray*}
\ell &:&W^{2,1}\X(M)\mapping\reals\\
\ell[u]&=&\int_M \left(|u|^2+|\grad u|^2\right)\d vol.
\end{eqnarray*}
Again, we consider the stationary points of $\ell$ with respect
to variations of the Euler-Poincar\'e form
\[
\delta{u}=\dot{\zeta}+[\zeta,u]
\,.
\]
This requirement yields the {\bf EPDiff equations}, for ``Euler-Poincar\'e
equations on the diffeomorphisms,'' given by the system, 
\begin{eqnarray}\label{EPDiff-eqn}
\dot{m}=-\mathrm{ad}_u^* m
\,,\quad\hbox{where}\quad
m=\frac{\delta \ell}{\delta u}
=u+\Delta^\grad u
\,.
\end{eqnarray}
Here we denote
\[
\Delta^\grad=\grad^*\grad
\]
for the connection Laplacian with respect to the metric and we
assume homogeneous boundary conditions.
\par
\paragraph{Diffeons: Singular solutions for EPDiff.}
The EPDiff equation (\ref{EPDiff-eqn}) has  Lie-Poisson Hamiltonian
structure in the momentum variable $m$ and satisfies a remarkable ansatz
for its singular solutions, which are called ``diffeons,''
\begin{eqnarray}\label{singsoln}
m(t,x) = \sum_{i=1}^N \int_{S_i}P_i(t,s_i)\delta(x-R_i(t,s_i)) \,ds_i
\,.
\end{eqnarray}
The diffeon singular solutions of EPDiff are vector-valued functions supported
in $M$ on a set of $N$ surfaces (or curves) of  codimension $(n-k)$ for
$s\in M^k\subset M$ with dim$M^k=k<n$.  For example, diffeons may be
supported on sets of points ($k=0$), curves ($k=1$), or surfaces ($k=2$)
in three dimensions. These support sets 
{of $m$} move with the fluid
velocity $u=G*m$; so the coordinates $s\in M^k\subset M$ are Lagrangian
fluid labels. The Green's function $G$ for the operator
$1+\Delta^\grad$ is continuous, but it has a jump in its derivative
on the support set that advects with the velocity $u=G*m$ under the
evolution of EPDiff. In fluid dynamics, a jump in derivative which moves
with the flow is called a ``contact discontinuity'' \cite{Le1992}. The
relationship of the singular solutions of EPDiff equation (\ref{EPDiff-eqn}) to
fluid dynamics is illuminated by rewriting equation (\ref{EPDiff-eqn}) in
``Riemann-invariant form,''
\[
\frac{d}{dt}(m\cdot{dx}\otimes{dvol})=0
\,,\quad\hbox{along}\quad
\frac{dx}{dt}=u=G*m
\,.
\]

The diffeon singular solution ansatz (\ref{singsoln}) was
first discovered as the ``peakon'' solutions for CH motion on the real
line in \cite{CaHo1993}. This was generalised to motion in higher
dimensions in \cite{HoSt2003} and was shown to be a momentum map in
\cite{HoMa2004}. As a result of the singular solution ansatz being a
momentum map, the $2N$ variables $P_i,R_i,$%
\rem{are canonically conjugate and they}
satisfy Hamilton's canonical equations. In general, these are integro-partial
differential equations for the canonically conjugate diffeon parameters
$P_i(t,s_i),R_i(t,s_i)$ in (\ref{singsoln}) with $i=1,\dots,N$.
\par
\subsection{Aim of the paper} This paper will be concerned with examining
diffeon solutions of the EPDiff equations (\ref{EPDiff-eqn}) on Einstein
manifolds that are ``mostly" symmetric, i.e., that have a group acting by
isometries so that the orbits have co-dimension 1 in the manifold except on a
set of measure zero, and hence the quotient of the manifold by the group is
1-dimensional.
\rem{In this situation,  the Green's function $G$ for the operator
$1+\Delta^\grad$ appearing in the EPDiff theory is continuous, but it has a
jump in its derivative that advects with the velocity $u=G*m$ under the
evolution of EPDiff.}
\par
Thus, we are interested in identifying and analysing
cases where imposing an additional translation symmetry on the solution
reduces the  canonical Hamiltonian dynamics of the singular solutions of
the EPDiff on Einstein manifolds from integro-partial differential
equations to ordinary differential equations for $P_i(t),R_i(t)$. We
shall then analyse some of the properties of those canonical equations and
examine their numerical solutions.
\par
We begin by noting some simplifications of the EPDiff equations when
restricted to Einstein spaces. These simplifications take advantage of the
relation between the connection Laplacian and the Hodge Laplacian, the latter
of which is easier to compute. Knowing the EPDiff equations in 1-dimension, we
move up to the Einstein spaces in two dimensions.
\par
The first manifold we study is the 2-sphere, which has the familiar group
action formed by rotations about a fixed axis. Thus we construct
``Puckons,'' which are singular solutions of the EPDiff equations
supported on concentric circular latitudes of the 2-sphere. The name ``Puckon''
(as opposed to ``peakon'') arises from the famous boast by Shakespeare's
character, Puck, in {\it A Midsummer Night's Dream}, that he would, ``put a
girdle round about the earth in forty minutes.''%
\footnote{Thanks to J. D. Gibbon for reminding us of this quote and suggesting
the name, ``Puckon'' for these solutions.}
 The case of rotationally symmetric singular solutions of EPDiff in the
Euclidean plane has already been discussed in \cite{HoPuSt2004}, so we proceed
to examine the hyperbolic plane, which has a rich isometry group. However, the
diffeon dynamics on the unbounded hyperbolic plane affords little opportunity
for multiple diffeon interactions and does not admit periodic behaviour. Richer
opportunities for diffeon dynamics and interactions in  hyperbolic spaces are
offered in the context of Teichm\"uller theory, which is, however, beyond the
scope of the present work. 
\par

From consideration of these 2-dimensional spaces, we sketch
how one might develop the theory for the general manifold which possesses a
metric equivalent to the warped product of the line with the bi-invariant
metric of a Lie group.

\subsection{Plan of the paper} After briefly recalling the essentials
needed from the theory of Einstein manifolds in section
\ref{ES-sec}, we begin in section \ref{Sphere-sec} by reducing the
singular solutions of EPDiff to a canonical Hamiltonian dynamical system
for the simplest Einstein manifold -- namely, motion on a sphere of
singular EPDiff solutions supported on concentric circular latitudes, or
``girdles.'' These new singular solutions girdling the sphere are the
``Puckons.''
Although the canonical reduction is {\it guaranteed} by the momentum map
property of EPDiff, the reduction of its singular diffeon solutions for Puckons
is given in detail, so it may be used to confirm the numerical results for the
interactions of Puckons on the sphere described in Section \ref{ES-sec}.
Section \ref{OtherSurfaces-sec} generalises the Puckons to other surfaces that
are Einstein manifolds with a translation symmetry. Section
\ref{warp-sec} incorporates these ideas into the theory of warped
products.

\section{The EPDiff Equations on Einstein Spaces}\label{ES-sec}
The operator
\[
\Delta^\grad+\id
\]
arises in the following context in Riemannian geometry.
\begin{defn1}
On any given Riemannian manifold, the musical isomorphisms are
defined to be the Riesz representation and inverse maps with
respect to the metric:
\begin{eqnarray*}
\sharp:\mathrm{T}^*M&\mapping &\mathrm{T}M\\
\inner{\alpha^\sharp}{w}&=&\alpha(w)\\
\\
\flat:\mathrm{T}M&\mapping &\mathrm{T}^*M\\
v_\flat(w)&=&\inner{v}{w}.
\end{eqnarray*}
\end{defn1}
In Riemannian geometry, the Levi-Civita
connection respects the musical isomorphisms, namely for
$v,w\in\X(M)$ and $\alpha\in\Form{M}{1}$:
\begin{eqnarray*}
(\grad_vw)_\flat&=&\grad_v(w_\flat)\\
(\grad_v\alpha)^\sharp&=&\grad_v(\alpha^\sharp).
\end{eqnarray*}
Thus, the musical isomorphisms identify vector fields with 1-forms and
allow them to be differentiated in the same way\par
The connection Laplacian on 1-forms satisfies the Bochner-Weitzenb\"ock
formula.
\begin{thm1}[Bochner-Weitzenb\"ock]\label{BWF}
On a Riemannian manifold with Levi-Civita connection $\grad$, any
1-form $\alpha$ satisfies
\[
\Delta^\d\alpha=\Delta^\grad\alpha+\mathrm{Ric}(\alpha)
\]
where $\Delta^\d=\d^*\d+\d\d^*$ is the Hodge Laplacian formed from
the exterior derivative $\d$ on forms and $\mathrm{Ric}$ is the
Ricci curvature operator.
\end{thm1}
Now we restrict our attention to special forms of manifold, namely
Einstein Manifolds.
\begin{defn1}
An Einstein manifold is a Riemannian Manifold which satisfies
\[
\mathrm{Ric}=k\id
\]
for some constant $k$.
\end{defn1}
On an Einstein manifold, one may scale the metric so
that $k$ can be replaced by $-1,0,1$ depending on whether $k$ is
negative, zero or positive respectively. Thus on an Einstein
manifold, the Bochner-Weitzenb\"ock formula becomes
\[
\Delta^\d=\Delta^\grad+\mathrm{sign}(k)\id.
\]
We restrict our study further to Einstein manifolds with positive
$k$ (we shall call these positive Einstein manifolds) and scale to
$k=1$. Thus we have
\[
\Delta^\d=\Delta^\grad+\id.
\]
The implication of this for the EPDiff Lagrangian on vector
fields is as follows, for homogeneous boundary conditions:
\begin{eqnarray*}
\ell[\bu]&=&\int_M\left(|\bu|^2+|\grad \bu|^2\right)\d vol\\
       &=&\int_M\left(|\bu_\flat|^2+|\grad \bu_\flat|^2\right)\d vol\\
       &=&\int_M\inner{\bu_\flat}{\bu_\flat+\Delta^\grad \bu_\flat}\d vol\\
       &=&\int_M\inner{\bu_\flat}{\Delta^\d \bu_\flat}\d vol\\
      \Big(&=&\int_M\left(|\d \bu_\flat|^2+|\d^* \bu_\flat|^2\right)\d
       vol\Big).
\end{eqnarray*}
Hence, finding the stationary points of Lagrangian $\ell$ with the usual
Euler-Poincar\'e constraints on the variations implies the EPDiff equation
system of equations (\ref{EPDiff-eqn}) in the form 
\begin{eqnarray}\label{CH1}
\dot{\bm}=-\mathrm{ad}_\bu^*\bm
\,,\quad\hbox{where}\quad
\bm=\Delta^\d \bu_\flat
\,.
\end{eqnarray}
These equations generalize the EPDiff equations to
Einstein manifolds.
\section{The EPDiff equations on the Sphere}\label{Sphere-sec}
In two dimensions, the only 
manifold with positive Einstein constant is the standard round sphere which we
shall regard as the Riemann sphere. We use stereographic
projections to identify the complex plane with the sphere whose
North Pole is removed. This is equivalent to putting the metric
\[
g=\frac{4}{(1+x^2+y^2)^2}\left(\d x^2+\d
y^2\right)=\frac{4}{(1+|z|^2)^2}\d z\d\bar{z}
\]
on the plane.
\subsection{Rotationally Invariant Solutions}
We shall examine vector-field solutions $\bu$ of EPDiff in (\ref{CH1}) with
\[
\int_{\sphere{2}}\inner{\Delta^{\d}\bu_\flat }{\beta}=\int_{C}\beta
\]
for any smooth 1-form $\beta$ where $C$ is a circle of latitude on
the sphere. The quantity $\bm=\Delta^{\d}\bu_\flat$ is a distribution,
defined by its integration against a smooth function. Thus, we seek
weak, or singular, solutions of EPDiff on the sphere.
 We change coordinates on the sphere to assist us in this search. The metric
on $\reals^2\backslash{0}$ in polar coordinates is $\d
r^2+r^2\d\theta$. Thus we can regard the sphere minus both poles
as $\reals^2\backslash 0$ with the metric
\[
g_{\,\mathbb{S}}=\rho^2\left(\d r^2+r^2\d\theta^2\right)
\,,\quad\hbox{where}\quad
\rho=\frac{2}{(1+r^2)}
\,.
\]
\paragraph{Green's function for the Helmholtz operator on the Riemann
sphere.} We now seek solutions to the equation
\begin{equation}\label{lap}
\Delta^{\d}\bu_\flat=\delta(r-R)(k_1\d r+k_2\d\theta)
\end{equation}
where $R,k_1,k_2$ are constants. Let us assume that the velocity
one-form $\bu_\flat$ is invariant under rotations of the sphere about the axis
joining North and South Poles and is radial, i.e. we must solve
\begin{eqnarray}
\bu_\flat&=&a(r)\d r,\label{lap2a}\\
\Delta^{\d}\bu_\flat&=&P\frac{\delta(r-R)}{r\rho(R)^2}\d r\label{lap2b}.
\end{eqnarray}
Let us first solve the Green's function equation
\[
\Delta^\d(G(r,R)\d r)=\frac{\delta(r-R)}{r}\d r
\,,
\]
for one recognises that the single diffeon velocity is proportional to the
Green's function, i.e.,
\[
a(r)=PG(r,R)\,.
\]
To solve explicitly for the Green's function in our present case,
we begin by recalling
\[
\Delta^{\d}G(r,R)\d
r=-\,\Diff{r}\left(\frac{(1+r^2)^2}{4r}\diff{(rG(r,R))}{r}\right)\d r\,.
\]
Integrating the Green's function equation using this expression
yields
\[
G(r,R)\d r
=\left(\frac{A_1}{r}+\frac{2A_2}{r(1+r^2)}+\frac{2}{r(1+\max(r,R)^2)}\right)\d
r\,,
\]
for constants $A_1,A_2$. In particular, we can remove the
singularities at $r=0$ and $r=\infty$ by setting $A_1=0$ and
$$A_2=-\,\frac{1}{R(1+R^2)}\,,$$ whence
\[
G(r,R)\d r =\frac{2\min(r,R)^2}{rR(1+r^2)(1+R^2)}\d r .
\]
Similarly, the solution to
\[
\Delta^\d \tilde{G}(r,R)\d\theta=R\delta(r-R)\d\theta
\]
for the angular diffeon velocity component is
\[
\tilde{G}(r,R)\d\theta
=\frac{2R\min(r,R)^2}{r(1+r^2)(1+R^2)}\d\theta=R^2G(r,R)\d\theta
\,.
\]
Both $G$ and $\tilde{G}$ are continuous over the sphere, but each
has a jump in derivative at $r=R$.

Thus, we have proved the
following.
\begin{prop1} [Radial Green's function on the sphere] The solution to
(\ref{lap2a},\ref{lap2b}) for the radial Green's function on the sphere is
\[
PG(r,R)\d r=\frac{2P\min(r,R)^2}{rR(1+r^2)(1+R^2)}\d r
\,.
\]
\end{prop1}

\paragraph{Solution ansatz for EPDiff on the sphere.}
Following \cite{HoSt2003}, we propose a solution ansatz for EPDiff
velocity on the sphere as the following superposition of Green's functions,
\begin{equation}\label{sph-ans-1form}
\eta=\sum_{i=1}^N P_iG_{R_i}\d r
\end{equation}
where we denote
\[
G_R(r)=RG(r,R).
\]
and $P_i$, $R_i$ are $2N$ functions of time. The corresponding vector
field dual to $\eta$ is
\begin{equation}\label{sph-ans-vec}
\bu=\sum_{i=1}^N\frac{1}{\rho^2}P_iG_{R_i}\partial_r
\end{equation}
We pair the arbitrary smooth vector field $w=f\partial_r+g\partial_\theta$ on
$\sphere{2}$ with the EPDiff equation using the solution ansatz given by
the one-form $\eta$ in equation (\ref{sph-ans-1form}). A direct calculation
yields the following equations for the diffeon
parameters,
\begin{prop1}[Diffeon parameter equations on the sphere]
\hfill\\ 
The equations for diffeon parameter evolution on the sphere are:
\vspace{-2mm}
\begin{eqnarray}
\dot{R_i}&=&\sum_{j=1}^N\left(\frac{P_j}
{\rho(R_i)^2\rho(R_j)^2R_j}G(R_i,R_j)\right)
\label{SCH1}\\
\dot{P}_i&=&-\sum_{j=1}^N\frac{P_iP_j}{\rho(R_i)^2\rho(R_j)^2}
\left(\Diff{r}G(r,R_j)\bigg\vert_{r=R_i}+2R_i\rho(R_i)G(R_i,R_j)\right)
\label{SCH2}.
\end{eqnarray}
\end{prop1}      
{\bf Proof (by direct calculation):}
\begin{eqnarray*}
0&=&\Diff{t}\left(\int_{\sphere{2}}\Delta^{\d}\eta(w)\d vol\right)
 +\int_{\sphere{2}}\Delta^{\d}\eta([w,\bu])\d vol\\
&=&\Diff{t}\left(\sum_{i=1}^N\int_0^{2\pi}\int_0^\infty\delta(r-R_i)f(r,\theta)P_i\rho^2r\d r\d \theta\right)\\
 &+&\int_0^{2\pi}\sum_{i=1}^N\int_0^\infty\delta(r-R_i)P_i\d
 r\left(\left[f(r,\theta)\partial_r+g(r,\theta)\partial_\theta,\sum_{j=1}^N{\rho^{-2}}P_jG_{R_j}(r)\partial_r\right]\right)\rho^2r\d r\d\theta\\
&=&\Diff{t}\left(\sum_{i=1}^N P_i\int_0^{2\pi}f(R_i,\theta)\d\theta\right)\\
 &+&\sum_{i,j=1}^NR_i\int_0^{2\pi}\left(P_jG_{R_j}'(R_i)f(R_i,\theta)+2P_jR_i\rho(R_i)G_{R_j}(R_i)f(R_i,\theta)
 -P_jG_{R_j}(R_i)f_r(R_i,\theta)\right)P_i\d\theta\\
&=&\sum_{i=1}^N \int_0^{2\pi}(f_r(R_i,\theta)\dot{R_i}P_i+\dot{P}_if(R_i,\theta))\d\theta\\
 &+&\sum_{i,j=1}^NR_i\int_0^{2\pi}\left(P_jG_{R_j}'(R_i)f(R_i,\theta)+2P_jR_i\rho(R_i)G_{R_j}(R_i)f(R_i,\theta)
 -P_jG_{R_j}(R_i)f_r(R_i,\theta)\right)P_i\d\theta\\
\end{eqnarray*}
Comparing coefficients of $f$ and $f_r$ implies
\begin{eqnarray*}
0&=&\dot{R_i}P_i-\sum_{j=1}^N\left(\frac{P_iP_j}{\rho(R_i)^2\rho(R_j)^2}G(R_i,R_j)\right)\\
0&=&\dot{P}_i+\sum_{j=1}^N\frac{P_iP_j}{\rho(R_i)^2\rho(R_j)^2R_j}\left(G_{R_j}'(R_i)+2R_i\rho(R_i)G_{R_j}(R_i)\right)\\
\end{eqnarray*}
This finishes the calculation of the diffeon parameter
evolution equations (\ref{SCH1},\ref{SCH2}). \hfill\fbox{$\surd$}
\begin{prop1}[Canonical Hamiltonian form of diffeon parameter
equations] \hfill\\
Evolution equations (\ref{SCH1},\ref{SCH2}) for the diffeon parameters are
equivalent to Hamilton's canonical equations,
\begin{eqnarray}\label{can-eqns-puckon}
\dot{R_i}=\frac{1}{\pi}\diff{H}{R_i}\,,\quad
\dot{P}_i=-\frac{1}{\pi}\diff{H}{P_i}\,,
\end{eqnarray}
with Hamiltonian function of $\mathbf{P}=(P_1,\ldots,P_N)$ and
$\mathbf{R}=(R_1,\ldots,P_N)$ given by,
\begin{eqnarray}\label{puckon-Ham}
H(\mathbf{P},\mathbf{R})
=\frac{\pi}{2}\sum_{i,j=1}^N
\frac{P_iP_jG(R_j,R_i)}{\rho(R_i)^2\rho(R_j)^2}
\,.
\end{eqnarray}
As explained in \cite{HoMa2004}, this reduction of EPDiff to canonical
Hamiltonian form for the diffeons is guaranteed, since the singular
solution ansatz (\ref{singsoln}) is a momentum map. However, we shall require
the explicit results for the numerical solutions in section
\ref{NumSol}.
\end{prop1}
{\bf Proof:} We Legendre transform the Lagrangian $\ell$ into the Hamiltonian
$H$ by setting
\[
H(\bm)=\int_{\sphere{2}}\bm(\bu)\d vol-\ell[\bu]
\]
whence we find that
\[
H(\bm)=\frac{1}{2}\int_{\sphere{2}}\bm(\bu)\d vol
\]
with
\[
\bm=\Delta^\d\bu_\flat.
\]
We then express the velocity as the superposition of Green's functions,
\[
\bu_\flat=\eta=\sum_{i=1}^N\frac{1}{\rho(R_i)^2}
\frac{2P_i\min(r,R_i)^2}{rR_i(1+r^2)(1+R_i^2)}\d r
\,.
\]
We set $k_i=P_i/(\rho(R_i)^2R_i)$, for $i=1,\dots,N$, and we evaluate the
Hamiltonian on this solution with 
$
\bm=\Delta^\d\eta
$
as
\begin{eqnarray*}
H(\bm)&=&\frac{1}{2}\int_{\sphere{2}}\inner{\eta}{\bm}\d vol\\
          &=&\frac{1}{2}\int_0^{2\pi}\int_0^\infty\sum_{i,j=1}^N\inner{\frac{2k_i\min(r,R_i)^2}{r(1+r^2)(1+R_i^2)}\d r}{\Delta^\d\frac{2k_j\min(r,R_j)^2}{r(1+r^2)(1+R_j^2)}\d r}\rho^2r\d
          r\wedge\d\theta\\
          &=&\frac{1}{2}\int_0^{2\pi}\int_0^\infty\sum_{i,j=1}^N\inner{k_iG_{R_i}(r)\d r}{k_j\delta(r-R_j)\d r}\rho^2r\d
          r\wedge\d\theta\\
          &=&\frac{1}{2}\int_0^{2\pi}\sum_{i,j=1}^N
          k_iG_{R_i}(R_j)k_jR_j\d\theta\\
          &=&\frac{\pi}{2}\sum_{i,j=1}^N\frac{P_iP_jG(R_i,R_j)}{\rho(R_i)^2\rho(R_j)^2}.
\end{eqnarray*}
The canonical equations for this Hamiltonian now recover
the diffeon parameter evolution equations (\ref{SCH1},\ref{SCH2}).
\hfill\fbox{$\surd$}

\begin{rmk1}
[Remark on constant factors] For convenience and brevity of
writing, in what follows we shall often absorb into the time units any constant
factors -- such as $\pi$ in equations (\ref{can-eqns-puckon}-\ref{puckon-Ham})
-- arising from integration over symmetry directions.
\end{rmk1}

\begin{defn1}[$N$-Puckon]
The singular solution of EPDiff 
\begin{eqnarray}
\Diff{t}\Delta^\d\eta&=&-\mathrm{ad}^*_{\eta^\sharp}\Delta^\d\eta
\,,
\end{eqnarray}
is given on the Riemann sphere by a vector field $\eta$ satisfying
\begin{eqnarray}\label{puckon-vel}
\Delta^\d\eta&=&\sum_{i=1}^N\frac{P_i}{\rho(R_i)^2R_i}\delta(r-R_i)\d r
\,.
\end{eqnarray}
The support set of this (weak) solution of EPDiff on the Riemann sphere is a set
of circular latitudes (girdles) at radii $R_i(t)$ with conjugate radial momenta
$P_i(t)$, where $i=1,\dots,N$. Equation (\ref{puckon-vel})
constitutes a solution ansatz for the velocity vector field $\eta$, which will
be called an
$N$-Puckon solution. 
\end{defn1}

\subsection{The Basic Irrotational Puckon}
\paragraph{Purpose.} This subsection derives explicit solutions for the
parameters $P$ and $R$ of the single Puckon without rotation and examines its
behaviour upon collapsing to the poles of the sphere.

\paragraph{The single irrotational Puckon.} Let us consider the case where
$N=1$ and examine the motion of the basic Puckon without rotation. For $N=1$,
the Hamiltonian (\ref{puckon-Ham}) is given by
\[
H(P,R)=\frac{1}{2}\frac{P^2G(R,R)}{\rho(R)^4}=\frac{P^2}{8}(1+R^2)^2
\,,
\]
which follows because the Green's function in this case is
\[
G(R,R)=2\frac{\min(R,R)^2}{R^2(1+R^2)^2}=\frac{2}{(1+R^2)^2}
\,.
\]
Thus, upon restricting ourselves to the constant level set of the
Hamiltonian defined by
\[
H(P,R)=\frac{K^2}{2}
\]
for constant $K$, we may solve for the momentum variable,
\[
P=K\rho(R)=\frac{2K}{(1+R^2)}
\,.
\]
Next, we note that the canonical coordinate equation
\[
\dot{R}=\diff{H}{P}=\frac{P}{4}(1+R^2)^2=\frac{K}{2}(1+R^2)
\,,
\]
integrates to 
\begin{equation}\label{R-eqn}
R=\tan\left(\frac{K}{2}t+\frac{\eps}{2}\right)
\,.
\end{equation}
Consequently, the single Puckon momentum is found as 
\begin{equation}\label{P-eqn}
P=K+K\cos\left({K}t+{\eps}\right).
\end{equation} 
 So far we have been using the stereographic projection
to provide charts for the sphere. However we may easily pass from
the coordinates $(r,\theta)$ obtained from stereographic
projection to latitudinal-longitudinal coordinates $(\phi,\theta)$
where $r$ and $\phi$ are related by
\[
r=\tan\left(\frac{\phi}{2}\right).
\]
By this token, the colatitude $\phi=\Phi(t)$ of the peak of the Puckon
evolves linearly in time as
\[
\Phi(t)=Kt+\eps.
\]
\begin{prop1}[Irrotational Puckon Solution]
The velocity vector field $v=\eta^\sharp$
generated by the irrotational Puckon motion is given in stereographic
coordinates by
\[
\bu=\frac{P}{4}\frac{\min(r,R)}{\max(r,R)}(1+r^2)\partial_r
\,.\]
This appears in latitudinal-longitudinal coordinates on the Riemann sphere, as
follows,
\[
\bu=\frac{K(1+\cos(Kt+\eps))}{2}
\frac{\min\left(\left|\tan\left(\frac{\phi}{2}\right)\right|,
\left|\tan\left(\frac{K}{2}t+\frac{\eps}{2}\right)\right|\right)}
{\max\left(\left|\tan\left(\frac{\phi}{2}\right)\right|,
\left|\tan\left(\frac{K}{2}t+\frac{\eps}{2}\right)\right|\right)}\Diff{\phi}.
\]
\end{prop1}
\paragraph{Proof:} This result is obtained by direct substitution of solutions
(\ref{R-eqn}) and (\ref{P-eqn}) for the diffeon parameters into the solution
ansatz (\ref{sph-ans-vec}).

\begin{rmk1} [Puckon peak]
The peak of the Puckon occurs when
\[
\phi=\Phi(t)=Kt+\eps,
\]
(modulo behaviour at the poles), whence
\[
|\bu|_{\sphere{2}}=\frac{|K|(1+\cos(Kt+\eps))}{2}
\,.
\]
\end{rmk1}

We still need to examine precisely what happens to the dynamics of
a Puckon as it collides with itself at the poles.

\begin{prop1} [Puckon behavior at the poles]
The Puckon bounces elastically, when it collides with itself at the
poles.
\end{prop1}

{\bf Proof:}
We notice that $\dot{R}=K/2\neq0$ at $R=0$ and by setting $U=1/R$ we see
that at $U=0$, $\dot{U}=K/2\neq0$. Similarly, one can show
that $\dot{P}=0$ at both poles. Thus, the Puckon velocity is
{\it finite} at the poles. Since Hamiltonian motion is time-reversible, the
Puckon must bounce elastically, when it collides with itself at the
poles.

\subsection{Rotating Puckons}
So far we have concentrated on singular EPDiff solutions on the sphere
moving with only a radial component. Now we turn to examine the rotating
$N$-Puckon, i.e. the solution of the following system of equations for $\eta$,
cf. equations (\ref{lap}),
\begin{eqnarray}
\label{rp1}\Delta^\d\eta&=&\sum_{i=1}^N\frac{\delta(r-R_i)}{\rho(R_i)^2R_i}
\left(P_i\d r+M_i\d\theta\right)\,,\\
\label{rp2}\Diff{t}\Delta^\d\eta&=&-\mathrm{ad}^*_{\eta^\sharp}\Delta^\d\eta
\,,
\end{eqnarray}
in which $M_i(t)$ with $i=1,\dots,N$, are the {\bf angular momenta} of the $N$
Puckons.

\begin{prop1}[Canonical Hamiltonian equations for the rotating $N$-Puckon]
\label{RotPuckEqns}\hfill\\
The equations for the rotating $N$-Puckon may be expressed as:
\begin{eqnarray}
\label{RCH1}
\dot{R}_i&=&\sum_{j=1}^NP_jF_j(R_i)R_jR_i
=\frac{\partial H}{\partial P_i}
\,,\\
\label{RCH2}
\dot{P}_i&=&-\sum_{j=1}^N
\left(P_iP_jF_j'(R_i)R_jR_i+P_iP_jF_j(R_i)R_j
+M_iM_jF_j'(R_i)\right)
=
-\,\frac{\partial H}{\partial R_i}
\,,\\
\label{RCH3}\dot{M}_i&=&0
\,,
\end{eqnarray}
where $F_i(R_j)=F_j(R_i)$ is defined in terms of the Green's function $G$
as
\[
F_i(R_j)=\frac{G(R_i,R_j)}{\rho(R_i)^2R_i\rho^2(R_j)R_j}
\,.
\]
The last equation (\ref{RCH3}) expresses the conservation of the angular
momentum of each Puckon. 
With $M_i$ constant, the first two equations (\ref{RCH1}) and (\ref{RCH2})
comprise Hamilton's canonical equations with symmetry-reduced Hamiltonian
\begin{eqnarray}\label{NPuck-Ham}
H(\mathbf{P},\mathbf{R};\mathbf{M})&=&\frac{1}{2}\sum_{i,j=1}^N
\left(P_iP_jR_iR_j+M_iM_j\right)F_i(R_j)\,.
\end{eqnarray}
\end{prop1}

\paragraph{Proof.}
First, we note that the velocity one-form corresponding to the momentum singular
solution ansatz (\ref{rp1}) is given by, 
\begin{eqnarray*}
\eta&=&\sum_{i=1}^N\frac{G(r,R_i)}{\rho(R_i)^2}\left(P_i\d
r+M_ir\d\theta\right)\\
\end{eqnarray*}
Thus the velocity vector field $\bu$ dual to $\eta$ is given by
\begin{eqnarray*}
\bu&=&\sum_{i=1}^N\frac{G(r,R_i)}{\rho(R_i)^2R_i\rho^2r}\left(R_iP_ir\partial_r+M_i\partial_\theta\right).
\end{eqnarray*}
To make our calculations more transparent, we collect terms and define
notation as
\[
F_i(r)=\frac{G(r,R_i)}{\rho(R_i)^2R_i\rho^2r},
\]
so that
\[
\bu=\sum_{i=1}^NF_i(r)\left(R_iP_ir\partial_r+M_i\partial_\theta\right)
\]
and
\[
\eta=\sum_{i=1}^NF_i(r)\rho^2r\left(R_iP_i\d
r+M_ir\d\theta\right).
\]
Note that the expression for $F_i$ is symmetric in $r$ and $R_i$, which
implies the permutation symmetry,
\[
F_i(R_j)=F_j(R_i).
\]
Now let $w=f\partial_r+g\partial_\theta$ be any smooth vector
field on $\sphere{2}$. Then, one computes the pairing
\begin{eqnarray*}
\int_{\sphere{2}}\Delta^\d\eta(w)\d vol
&=&\sum_{i=1}^N\int_0^{2\pi}\left(P_if(R_i,\theta)+M_ig(R_i,\theta)\right)\d\theta.
\end{eqnarray*}
Hence, the left hand side of the EPDiff equation becomes
\[
\Diff{t}\left(\int_{\sphere{2}}\Delta^\d\eta(w)\d vol\right)
=\sum_{i=1}^N\int_0^{2\pi}
\left(\dot{P}_if(R_i,\theta)+P_i\dot{R}_if_r(R_i,\theta)+\dot{M}_ig(R_i,\theta)+M_i\dot{R}_ig_r(R_i,\theta)\right)\d\theta.
\]
Now we need to calculate its right hand side.
\[
\int_{\sphere{2}}\Delta^\d\eta([w,\bu])\d vol.
\]
For this, we write the commutator,
\begin{eqnarray*}
[w,\bu]&=&\sum_{i=1}^N\left(F_i'R_iP_irf+F_iR_iP_if-F_iR_iP_irf_r-F_iM_if_\theta\right)\partial_r\\
     &+&\sum_{i=1}^N\left(fF_i'M_i-F_iR_iP_irg_r-F_iM_ig_\theta\right)\partial_\theta.
\end{eqnarray*}
Now, by Stokes' theorem, when we integrate over the whole sphere,
the contribution due to $f_\theta,g_\theta$ will be zero because
$f,g$, the only components of the integration to depend on
$\theta$, satisfy  $f(r,0)=f(r,2\pi)$ etc. Thus
\begin{eqnarray*}
& &\int_{\sphere{2}}\Delta^\d\eta([w,\bu])\d vol\\
&=&\sum_{i,j=1}^N\int_0^{2\pi}
\left(P_iF_j'(R_i)R_jP_jR_if(R_i,\theta)+P_iF_j(R_i)R_jP_jf(R_i,\theta)-P_iF_j(R_i)R_jP_jR_if_r(R_i,\theta)\right)\d\theta\\
&&+\sum_{i,j=1}^N\int_0^{2\pi}\left(M_iF_j'(R_i)M_jf(R_i,\theta)-M_iF_j(R_i)R_jP_jR_ig_r(R_i,\theta)\right)\d\theta.
\end{eqnarray*}
Now to find the evolution equations for $P_i,R_i,M_i$ we need only 
compare the coefficients of $f,f_r,g,g_r$ occurring in
\[
0=\Diff{t}\left(\int_{\sphere{2}}\Delta^\d\eta(w)\d
vol\right)+\int_{\sphere{2}}\Delta^\d\eta([w,\bu])\d vol
\,.
\]
From this comparison, one finds
\begin{eqnarray*}
0&=&\dot{P}_i
+\sum_{j=1}^N\Big(P_iF_j'(R_i)R_jP_jR_i+P_iF_j(R_i)R_jP_j
+M_iF_j'(R_i)M_j\Big)\,,\\
0&=&P_i\dot{R}_i -\sum_{j=1}^n\Big(P_iF_j(R_i)R_jP_jR_i
\Big)\,,\\
0&=&\dot{M}_i\,,\\
0&=&M_i\dot{R}_i-\sum_{j=1}^N\Big(M_iF_j(R_i)R_jP_jR_i\Big)\,,
\end{eqnarray*}
in which the second and fourth equations provide the {\it same} information.
We therefore obtain the evolution equations (\ref{RCH1}-\ref{RCH3}) for
$P_i,R_i,M_i$ given in the statement of Proposition \ref{RotPuckEqns}.
\hfill\fbox{$\surd$}

\begin{rmk1} [Verifying the Hamiltonian]
A simple check shows that if the Hamiltonian
\[
H(\bm)=\int_{\sphere{2}}\bm(\bu)\d vol-\ell[\bu]
\]
is the Legendre transform of the Lagrangian $\ell$ then
\begin{eqnarray*}
H(\bm)&=&\frac{1}{2}\int_{\sphere{2}}\inner{\Delta^\d\eta}{\eta}\d vol\\
&=&2\pi H(\mathbf{P},\mathbf{R};\mathbf{M})
\,,
\end{eqnarray*}
which is the Hamiltonian for the rotating $N$-Puckon in formula
(\ref{NPuck-Ham}) of Proposition \ref{RotPuckEqns}. As explained in
\cite{HoMa2004}, the reduction to canonical Hamiltonian form is guaranteed,
since the singular solution ansatz (\ref{singsoln}) is a momentum map.
\end{rmk1}

\subsection{The Basic Rotating Puckon}\label{BasicRot-subsec}

\begin{prop1}
[Extremal radii of the basic rotating Puckon]
\hfill\\
The motion of the basic rotating Puckon lies between the maximum and minimum
values of $R$ on the Riemann plane given by,
\begin{eqnarray}\label{Rmax}
R_{max/min}&=&\frac{2K\pm \sqrt{4K^2-M^2}}{M}
\,,
\end{eqnarray}
assuming that $2K\ge{M}>0$.
\end{prop1}

\paragraph{Proof:}
We $N=1$ in Hamiltonian (\ref{NPuck-Ham}) to find
\[
H(P,R)=\frac{1}{16R^2}(1+R^2)^2(P^2R^2+M^2).
\]
We notice that for a rotating Puckon with $H=K^2=const$, the
girdle of the Puckon cannot have zero radius unless $M=0$, in
which case we return to the irrotational Puckon. Likewise, the
girdle radius cannot be infinite unless again $M=0$. Thus, a
rotating Puckon with $M\ne0$ is constrained to lie between a maximum and a 
minimum radius. To find these extremal radii, we must solve
\begin{eqnarray*}
\diff{H}{P}=\frac{P}{8}(1+R^2)^2&=&0\,,\\
H=\frac{1}{16R^2}(1+R^2)^2(P^2R^2+M^2)&=&K^2.
\end{eqnarray*}
The first equation can only be solved by $P=0$; for which the second becomes
\[
\frac{M^2}{16R^2}(1+R^2)^2=K^2.
\]
Assuming that $K$ and $M$ are both positive, we find that the
maximum and minimum values of $R$ are the roots,
\begin{eqnarray*}
R_{max/min}&=&\frac{2K\pm \sqrt{4K^2-M^2}}{M}
\end{eqnarray*}
This proves the proposition.\hfill\fbox{$\surd$}

\begin{rmk1} [A single critical point]
We observe that if $M=2K$, then the Puckon is radially static at the
equator and $P$ is identically zero.
The solution $(P,R)=(0,1)$ is the only critical point of the
Hamiltonian $H$ unless $M=0$ in which case the set defined by
$P=0$ is the critical manifold.
\end{rmk1}

\begin{prop1}
[Periodic motion of the basic rotating Puckon]
\hfill\\
The motion of the rotating Puckon is periodic, with period $2\pi/\sqrt{2H}$
determined by  the constant value of the Hamiltonian, $H$.
\end{prop1}
\paragraph{Proof using the colatitude representation of periodic rotating Puckon
motion:} To investigate the motion of the rotating Puckon, we pass
from stereographic coordinates to longitude-colatitude coordinates in
which we put
\[
R=\tan\frac{\Phi}{2}.
\]
As we will see later in the general case, this does not alter the
situation a great deal. For example, the momentum is given later in equation
(\ref{Moment0}), after substituting $\sin$ for $\psi$. The Hamiltonian for the rotating Puckon on the
sphere in longitude-colatitude coordinates
is
\begin{equation}\label{colong-lat}
H(P,\Phi)=\frac{1}{2}\left(P^2+\frac{M^2}{\sin^2\Phi}\right),
\end{equation}
in which $\Phi$ and $P$ are canonically conjugate variables.
Along the motion of the Puckon, $H$ is constant, say $H=K^2$. Thus,
taking the positive branch for $P$
\[
P=\sqrt{2K^2-\frac{M^2}{\sin^2\Phi}}
\]
yields the equation for the colatitude
\[
\dot{\Phi}=P=\sqrt{2K^2-\frac{M^2}{\sin^2\Phi}}.
\]
Integration of this ODE implies that
\begin{eqnarray*}
t&=&\int_{\Phi(0)}^{\Phi(t)}
\frac{\sin \Phi\,\d \Phi}{\sqrt{2K^2\sin^2\Phi-M^2}}\\
 &=& \frac{1}{K\sqrt{2}}
\left(
\arccos\left(\frac{K\sqrt{2}\cos \Phi(t)}{\sqrt{2K^2-M^2}}\right)
-
\arccos\left(\frac{K\sqrt{2}\cos \Phi(0)}{\sqrt{2K^2-M^2}}\right)
\right).
\end{eqnarray*}
Thus the colatitude for the single rotating Puckon is given as a
function of time by
\[
\Phi(t)=\sqrt{\frac{2K^2\cos^2\Phi(0)}{2K^2-M^2}}\cos(Kt\sqrt{2})
-\sqrt{\frac{2K^2\sin^2\Phi(0)-M^2}{2K^2-M^2}}\sin(Kt\sqrt{2}).
\]
These rotating Puckon solutions are periodic with period $2\pi/(K\sqrt{2})$,
determined by the constant value $K^2$ of the Hamiltonian.
\hfill\fbox{$\surd$}

\subsection{Puckons and Geodesics}

\begin{prop1}
[Geodesic motion of a point on the girdle of the rotating Puckon]
\hfill\\
A point on the girdle of the rotating Puckon moves along a
great circle at constant speed equal to $K\sqrt{2}$ determined by the value
$H=K^2$ of the Hamiltonian, $H$. The normal to its plane of motion is inclined
to the South Pole at the angle $\pi-\arctan(2R_{max})$ with
$R_{max}$ given in (\ref{Rmax}).
\end{prop1}

\paragraph{Proof:} For a rotationally symmetric surface $\mathcal{M}$ with
metric
\[
g=\d r^2+\psi(r)^2\d\theta,
\]
the equations for a curve $\bx=(R(t),\Theta(t))$ to be a geodesic
are equivalent to
\begin{eqnarray}
\label{Unitnorm} \dot{R}^2+\psi^2 \dot{\Theta}^2&=&1\\
\label{Claireaux}\psi^2 \dot{\Theta}&=&const.
\end{eqnarray}
So in the case of the sphere with metric $g=\d \phi^2+\sin^2\phi\,\d\theta$, 
if we define
\begin{equation}\label{thetadot}
\dot{\Theta}=u_\theta(\Phi)=\frac{M^2}{\sin^2\Phi}
\end{equation}
then we obtain a curve on the sphere with
coordinates $\bx=(\Phi,\Theta)$ and tangent vector
\[
\dot{\bx}(t)=(\dot{\Phi}\,,\,\dot{\Theta}).
\]
Now, the speed is given by 
\begin{eqnarray*}
|\dot{\bx}|^2&=&g(\dot{\bx},\dot{\bx})\\
               &=&\dot{\Phi}^2+(\sin^2 \Phi)\, \dot{\Theta}^2\\
               &=&P^2+\frac{M^2}{\sin^2\Phi}\\
               &=&2H(P,\Phi)=constant=2K^2.
\end{eqnarray*}
Thus a point on the geodesic has constant speed equal to $\sqrt{2H}$ and by
(\ref{thetadot}) we have
\[
(\sin^2\Phi)\dot{\Theta}=M^2=constant.
\]
Consequently, $\boldsymbol{\dot{\bx}}/(K\sqrt{2})$ satisfies
equations (\ref{Unitnorm}) and (\ref{Claireaux}), thereby determining a
geodesic on the sphere. This means that a point on the girdle of the Puckon
moves along a great circle at constant speed equal to $\sqrt{2H}$ and the
normal to its plane of motion is inclined to the South Pole at the angle
$\pi-\arctan(2R_{max})$, with $R_{max}$ given in equation 
(\ref{Rmax}).\hfill\fbox{$\surd$}

\subsection{Further Hamiltonian Aspects of Radial Solutions of EPDiff on
the Riemann Sphere} 

\begin{prop1}[Lie-Poisson Hamiltonian form of EPDiff on the Riemann
sphere]\label{LPform}
The radially symmetric solutions of EPDiff on the Riemann
sphere may be written  as
\begin{equation}\label{LP-form}
\Diff{t}\left(\begin{array}{c}m_r\\
rm_\theta\end{array}\right)=\mathcal{D}\left(\begin{array}{c}u_r\\
\frac{u_\theta}{r}\end{array}\right)
=\mathcal{D}\left(\begin{array}{c}\delta H/\delta m_r\\
\delta H/\delta (rm_\theta)\end{array}\right)
\,,
\end{equation}
where $\mathcal{D}$ is the skew-symmetric Hamiltonian operator given by
\begin{equation}\label{Ham-op}
\mathcal{D}=
\left(\begin{array}{cc}-\left({m_r\partial_r}+{\partial_r m_r}
                       -2{m_r\rho r}+\frac{m_r}{r}
                      \right) & -{rm_\theta \partial_r}\\
 \left(-{\partial_rrm_\theta}
                       +2{m_\theta}{r^2\rho}-{m_\theta}\right)&0
                       \end{array}\right)
\,,
\end{equation}
and the velocities $u_r$ and $u_\theta/r$ are given by the
variational derivatives of the Hamiltonian,
\[
u_r=\frac{\delta H}{\delta m_r}
\,,\quad
\frac{u_\theta}{r}=\frac{\delta H}{\delta rm_\theta}
\,.
\]
Equations (\ref{LP-form},\ref{Ham-op}) provide the Lie-Poisson
Hamiltonian form of the EPDiff equation for radially symmetric dynamics
on the Riemann sphere.
\end{prop1}

\paragraph{Proof:}
In the case that
\begin{eqnarray*}
\bu&=&\sum_{i=1}^N\frac{G(r,R_i)}{\rho(R_i)^2R_i\rho^2r}\left(R_iP_ir\partial_r+M_i\partial_\theta\right)\\
   &=:&{u_r}\partial_r+\frac{u_\theta}{r}\partial_\theta.
\end{eqnarray*}
where the $P_i$ and $R_i$ satisfy (\ref{RCH1}) and (\ref{RCH2}),
the associated momentum density is
\[
\bm=\Delta^{\d}\bu_\flat=m_r\d r+rm_\theta\d\theta,
\]
where
\begin{eqnarray*}
m_r&=&\sum_{i=1}^N\frac{\delta(r-R_i)}{\rho(R_i)^2R_i}P_i
\,,\quad\hbox{and}\quad
rm_\theta=\sum_{i=1}^N\frac{\delta(r-R_i)}{\rho(R_i)^2R_i}M_i.
\end{eqnarray*}
Let $w_1=\frac{f}{\rho^2r}\partial_r$ and
$w_2=\frac{g}{\rho^2r}\partial_\theta$ be two vector fields on
$\sphere{2}$. These satisfy the commutator relations,
\begin{eqnarray*}
\left[w_1,\bu\right]&=&\left[\frac{f}{\rho^2r}\partial_r,u_r\partial_r
                       +\frac{u_\theta}{r}\partial_\theta\right]\\
                    &=&\left(\frac{fu_r'}{\rho^2r}-\frac{u_r\partial_rf}{\rho^2r}
                       -2\frac{fu_r}{\rho}+\frac{fu_r}{\rho^2r^2}
                       -\frac{u_\theta\partial_\theta f}{\rho^2r^2}\right)\partial_r
                       +\left(\frac{fu_\theta'}{\rho^2r^2}-\frac{fu_\theta}{\rho^2r^3}\right)\partial_\theta,\\
\end{eqnarray*}
and
\begin{eqnarray*}
\left[w_2,\bu\right]&=&\left[\frac{g}{\rho^2r}\partial_\theta,u_r\partial_r
                       +\frac{u_\theta}{r}\partial_\theta\right]\\
                    &=&-\left(\frac{u_r\partial_rg}{\rho^2r}
                       +2\frac{gu_r}{\rho}-\frac{gu_r}{\rho^2r^2}+\frac{u_\theta\partial_\theta
                       g}{\rho^2r^2}\right)\partial_\theta.
\end{eqnarray*}
Thus, the EPDiff equations become, for $w_1$,
\begin{eqnarray*}
0&=&\int_\sphere{2}\left(\Diff{t}\bm(w_1)+\bm([w_1,\bu])\right)\d
vol\\
 &=&\int_0^{2\pi}\int_0^\infty\left(\diff{m_r}{t}\frac{f}{\rho^2r}
                +m_r\left(\frac{fu_r'}{\rho^2r}-\frac{u_r\partial_rf}{\rho^2r}
                       -2\frac{fu_r}{\rho}+\frac{fu_r}{\rho^2r^2}
                       -\frac{u_\theta\partial_\theta
                       f}{\rho^2r^2}\right)\right)\rho^2r\d r\d\theta\\
 &+&\int_0^{2\pi}\int_0^\infty
 rm_\theta\left(\frac{fu_\theta'}{\rho^2r^2}
     -\frac{fu_\theta}{\rho^2r^3}\right)\rho^2r\d
 r\d\theta\\
 &=&\int_0^{2\pi}\int_0^\infty\left(\diff{m_r}{t}{f}+m_r\left({fu_r'}-{u_r\partial_rf}
                       -2{fu_r\rho r}+\frac{fu_r}{r}
                      \right)\right)\d r\d\theta\\
 &+&\int_0^{2\pi}\int_0^\infty r^2m_\theta\left(\frac{fu_\theta'}{r^2}-\frac{fu_\theta}{r^3}\right)\d
 r\d\theta
\end{eqnarray*}
From this we obtain the radial equation
\begin{eqnarray}\label{LP-r}
\diff{m_r}{t}&=&-\left({m_r\partial_r}+{\partial_r m_r}
                       -2{m_r\rho r}+\frac{m_r}{r}
                      \right)u_r
 -{rm_\theta \partial_r}\frac{u_\theta}{r}.
\end{eqnarray}
Similarly, for $w_2$,
\begin{eqnarray*}
0&=&\int_\sphere{2}\left(\Diff{t}\bm(w_2)+\bm([w_2,\bu])\right)\d
vol\\
&=&\int_0^{2\pi}\int_0^\infty\left(\diff{rm_\theta}{t}\frac{g}{\rho^2r}-rm_\theta\left(\frac{u_r\partial_rg}{\rho^2r}
                       +2\frac{gu_r}{\rho}-\frac{gu_r}{\rho^2r^2}+\frac{u_\theta\partial_\theta
                       g}{\rho^2r^2}\right)\right)\rho^2r\d
 r\d\theta\\
 &=&\int_0^{2\pi}\int_0^\infty\left(\diff{rm_\theta}{t}{g}-rm_\theta\left({u_r\partial_rg}
                       +2{gu_r}{r\rho}-\frac{gu_r}{r}\right)\right)\d
 r\d\theta\\
\end{eqnarray*}
Hence we have the azimuthal equation
\begin{eqnarray}\label{LP-theta}
\diff{rm_\theta}{t}&=&\left(-{\partial_rrm_\theta}
                       +2{m_\theta}{r^2\rho}-{m_\theta}\right){u_r}.
\end{eqnarray}
Equations (\ref{LP-r}) and (\ref{LP-theta}) provide the Lie-Poisson form of
EPDiff on the Riemann Sphere. \hfill\fbox{$\surd$}

\section{Numerical Solutions for EPDiff on the Sphere}\label{NumSol}
\subsection{Overview}
We present numerical solutions to both the EPDiff partial differential equations
(\ref{LP-form},\ref{Ham-op}) and the corresponding ordinary differential
equations (\ref{RCH1},\ref{RCH2}) for Puckons.  Instead of using the
stereographic projection, the numerical solutions were calculated on the sphere
with colatitude-longitude $(\phi,\theta)$ coordinates
(and canonical variables $\Phi,P$ instead of $R,P$).  
The equations in $(\phi,\theta)$ coordinates on the sphere are obtained from
those in Section \ref{OtherSurfaces-sec} using $\psi=\sin$, so that
$g=\d\phi^2+\sin^2\phi\,\d\theta^2$. \par
We also introduce a length
scale $\alpha$ for our numerical solutions by effectively changing the
radius of the sphere from 1 to $\alpha$.
In this case $\alpha^2\Delta^\d=1+\alpha^2\Delta^\grad$
so that the Green's function is
\[
G(\phi,\Phi)=\frac\alpha2\left\{
  \begin{array}{ll}
    \left(\tan\frac\phi2\cot\frac\Phi2\right)^{1/\alpha},&\phi<\Phi \\
    \left(\tan\frac\Phi2\cot\frac\phi2\right)^{1/\alpha},&\phi>\Phi
  \end{array}
  \right..
\]

\subsection{Numerical specifications.}
Numerical simulations for diffeons on the sphere were performed using both PDEs
and ODEs. In simulating the Lie-Poisson partial differential equations
(\ref{LP-form},\ref{Ham-op}) in colatitude-longitude coordinates in Proposition
\ref{LPform}, fourth-order finite differences were used to calculate spatial
derivatives, and the momentum was advanced in time using a fourth-order
Runge-Kutta scheme.  The Hamiltonian was verified to be conserved to within
0.01\% of its initial value for the simulations with smooth initial velocity
distributions and to within 5-10\% for the  simulations when using one or two
Puckons as the initial velocity distribution.
The ordinary differential equations (\ref{RCH1},\ref{RCH2}) for the canonical
variables in colatitude-longitude coordinates were advanced in time using a 
fourth-order Runge-Kutta scheme.  The Hamiltonian was conserved for these ODE 
simulations to within $10^{-6}\%$.  For PDE simulations, the results will be
shown as a velocity  distribution changing in time,
whereas the results of the ODE simulations will be shown as plots of the time
evolution of the canonical variables $\Phi,P$.
The length scale $\alpha$ was set to $0.1$ unless otherwise noted.

\begin{figure}
\centering
\includegraphics[scale=0.5]{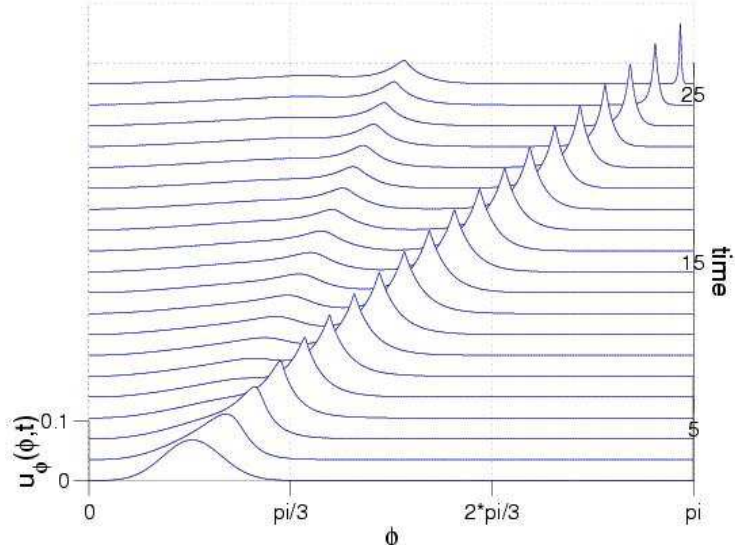}
\caption{An initial Gaussian breaks up into Puckons, two of which
     are evident for the times shown.}
\label{irrotIVP-fig}
\end{figure}

\paragraph{Irrotational Puckons.}
We first consider irrotational Puckons ($u_\theta=0)$.  
Figure \ref{irrotIVP-fig}
shows the evolution when the initial meridional velocity
$u_\phi(\theta,0)$ is a Gaussian.  
As time elapses ($t=0$ is at the 
bottom of the figure, and later times are shown above it),
a Puckon emerges from the Gaussian in Figure \ref{irrotIVP-fig}, and a
second Puckon begins to emerge.  In agreement with the prediction of 
equation (\ref{colong-lat}), the first Puckon retains its height (and thus
retains its velocity and canonical momentum $P$) as it approaches the South
Pole.

\begin{figure}
\centering
\includegraphics[scale=0.6]{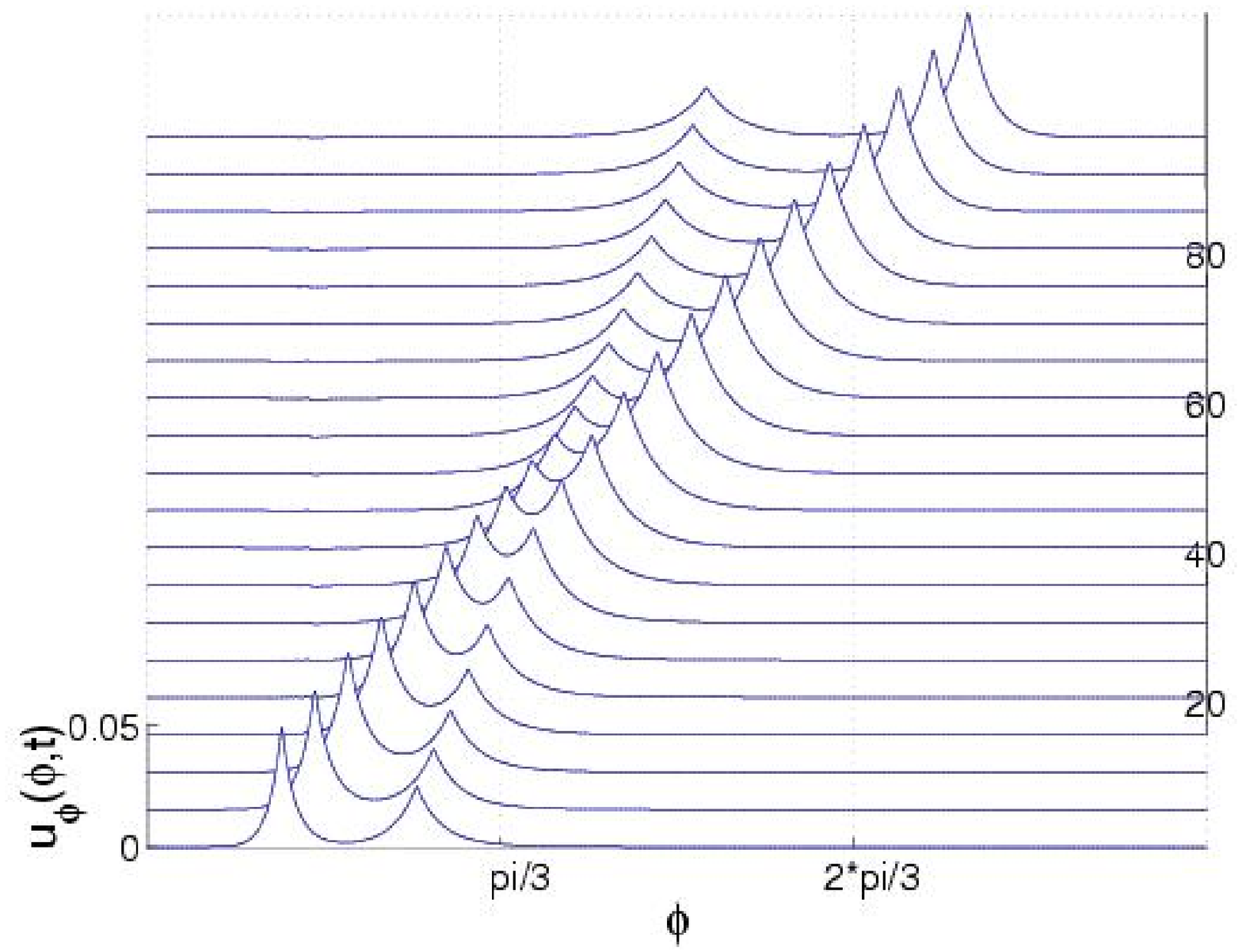}

\includegraphics[scale=0.4]{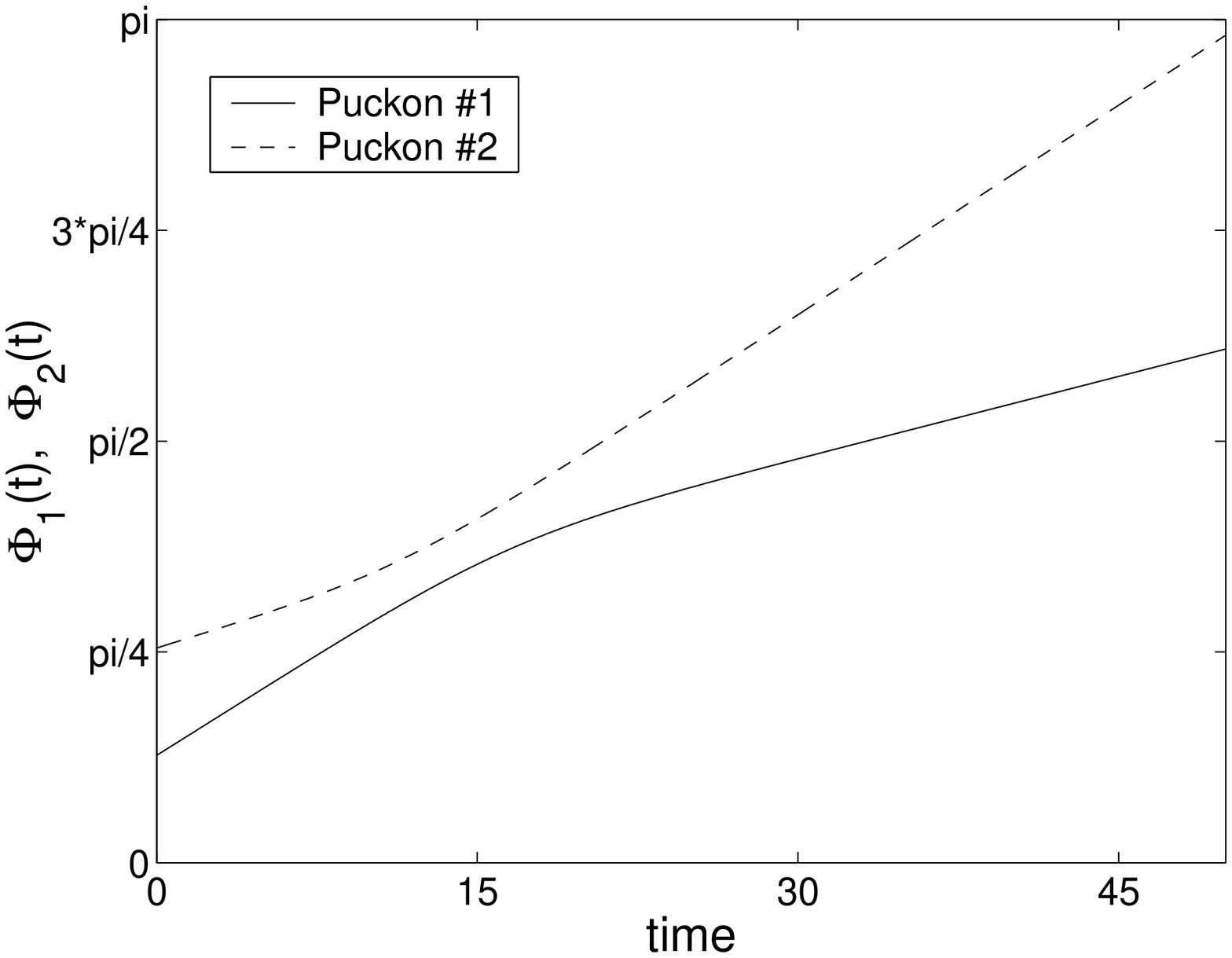}
\includegraphics[scale=0.4]{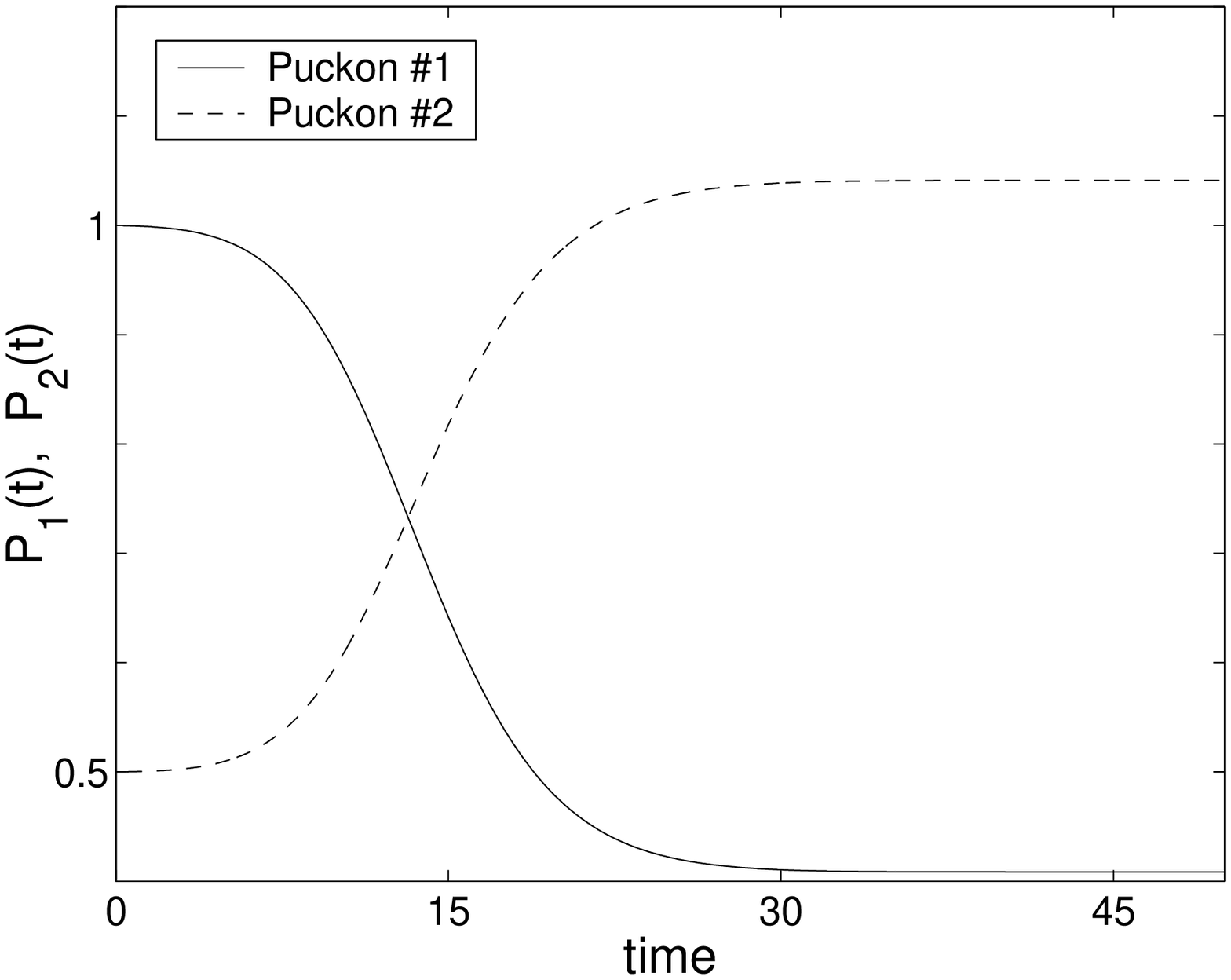}
\caption{Two Puckons undergo a rear-end collision.  The PDE
  simulation agrees with the colatitudes $\Phi$ of the ODE
  simulation to within 1\% and with the canonical momenta
  $P$ to within 3\%.}
\label{rear-fig}
\end{figure}

A rear-end collision between two irrotational Puckons is shown in 
Figure \ref{rear-fig}.
As the plot of colatitudes shows, these two Puckons do not pass through 
each other.  Instead, they bounce and exchange momenta. However,
their momenta are not exchanged exactly, as the plot of $P(t)$
shows.  This contrasts with 2-soliton collisions for
completely integrable PDEs in which momentum is exactly exchanged.

\begin{figure}
\centering
\includegraphics[scale=0.6]{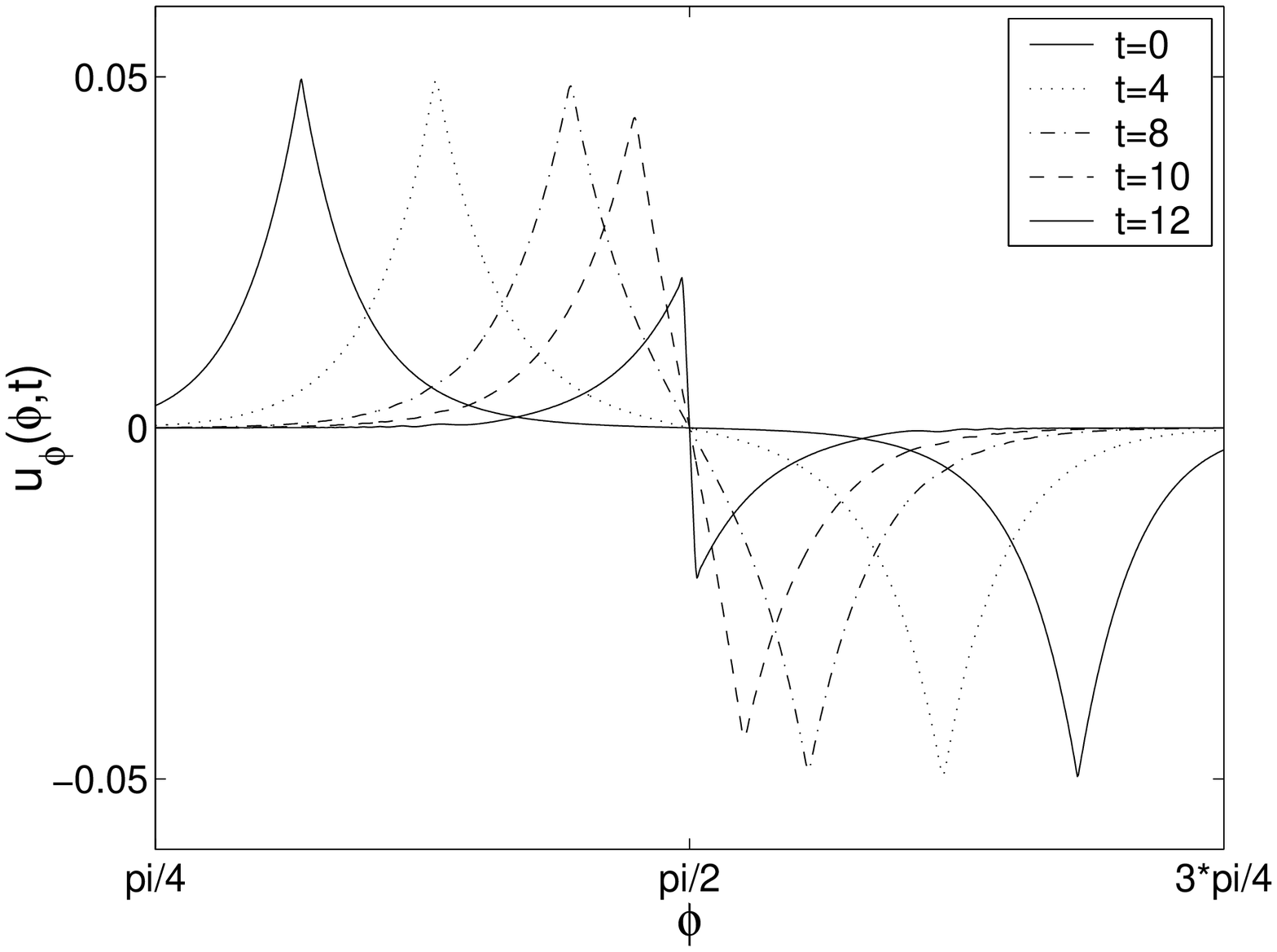}

\includegraphics[scale=0.4]{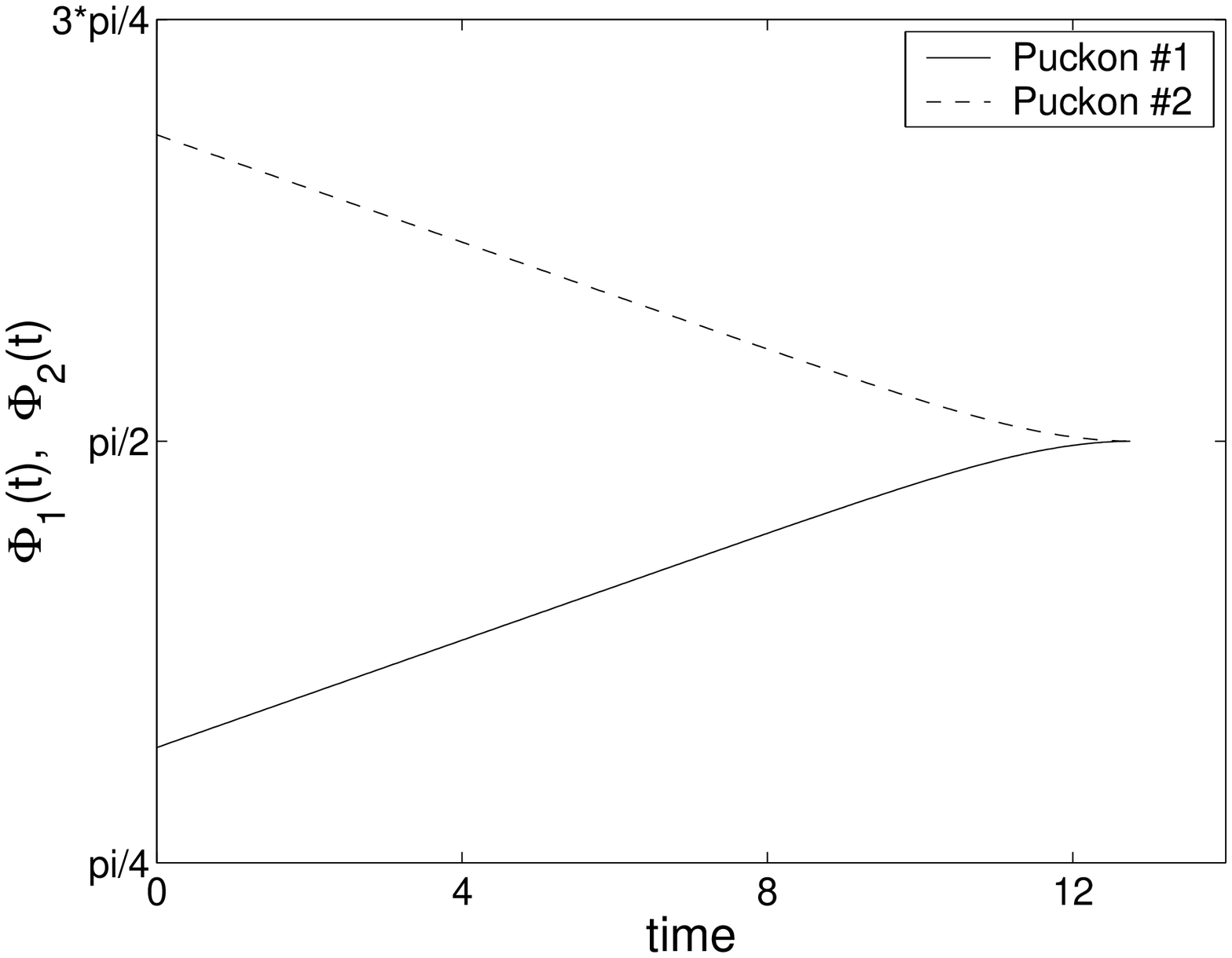}
\includegraphics[scale=0.4]{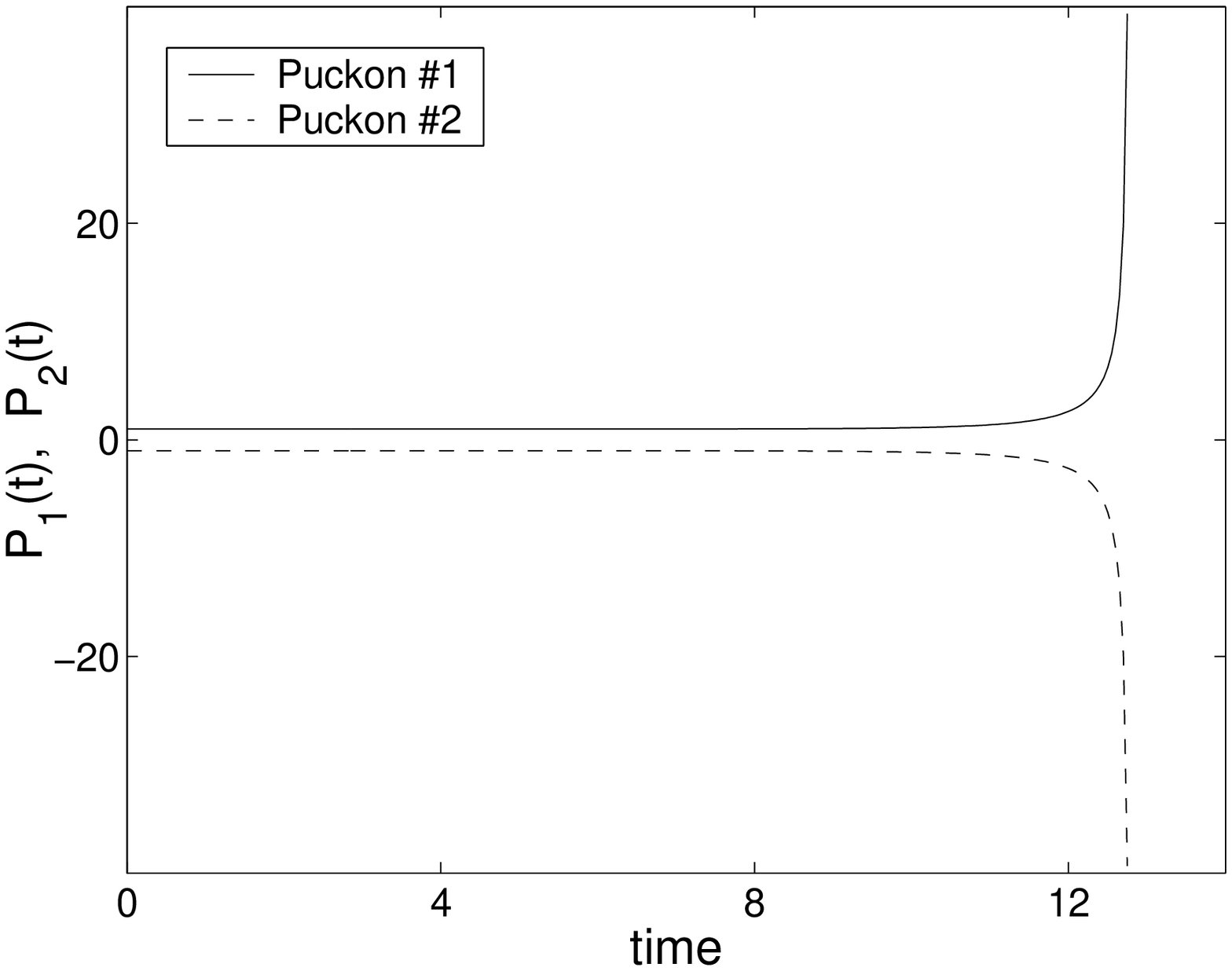}
\caption{Two Puckons undergo a head-on collision.}
\label{anti-fig}
\end{figure}

A head-on collision between two irrotational Puckons is shown in 
Figure \ref{anti-fig}.
In the PDE simulation, a vertical slope appears to form in finite time.
Note that the Puckon velocities (i.e., the Puckon heights in the
PDE simulation or the slopes of $\Phi(t)$ in the ODE simulation)
remain finite and actually decrease to zero, whereas the
equal and opposite canonical radial Puckon momenta  diverge as the collision
takes place.\par

\begin{figure}
\centering
\includegraphics[scale=0.6]{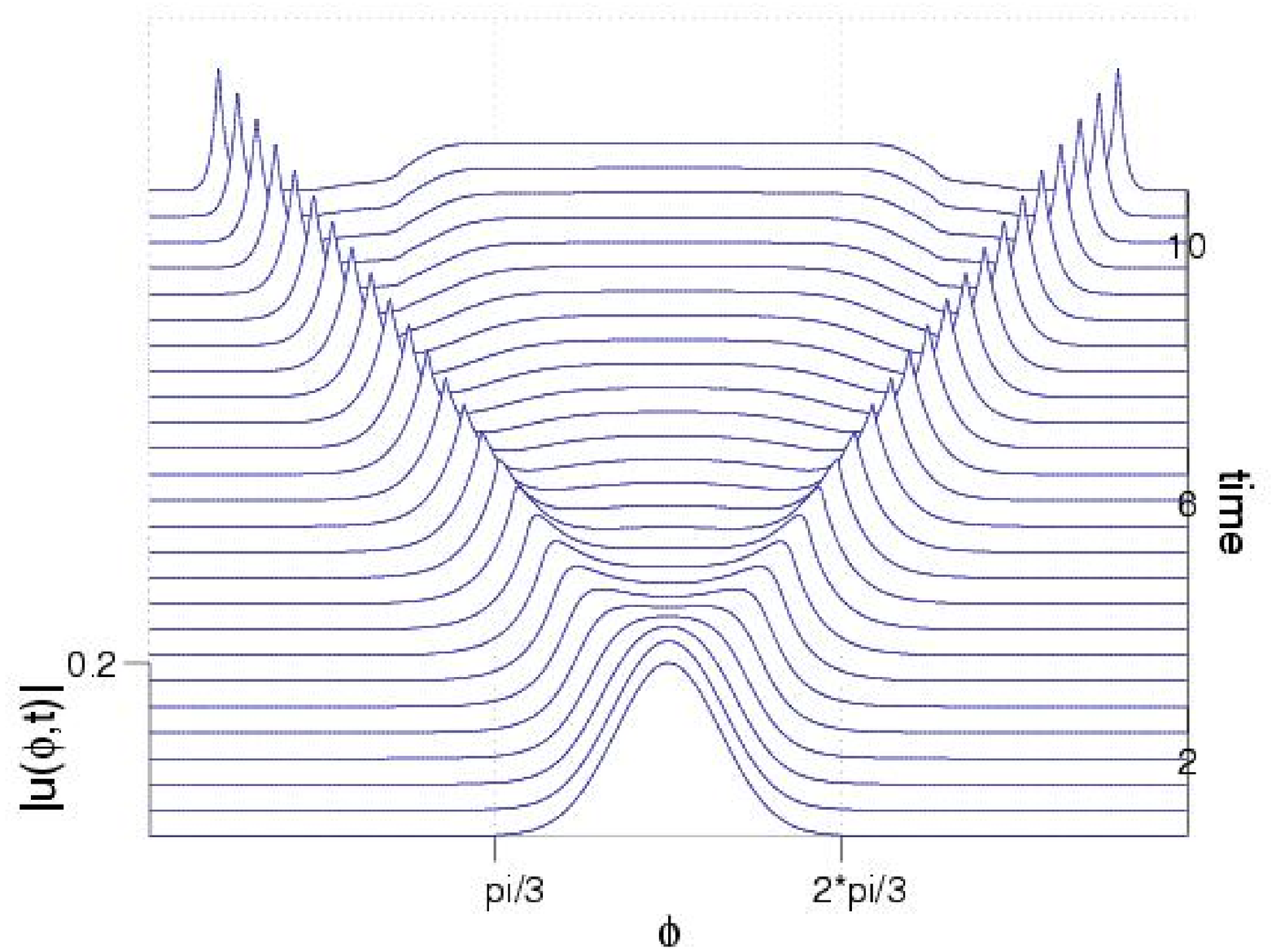}

\includegraphics[scale=0.4]{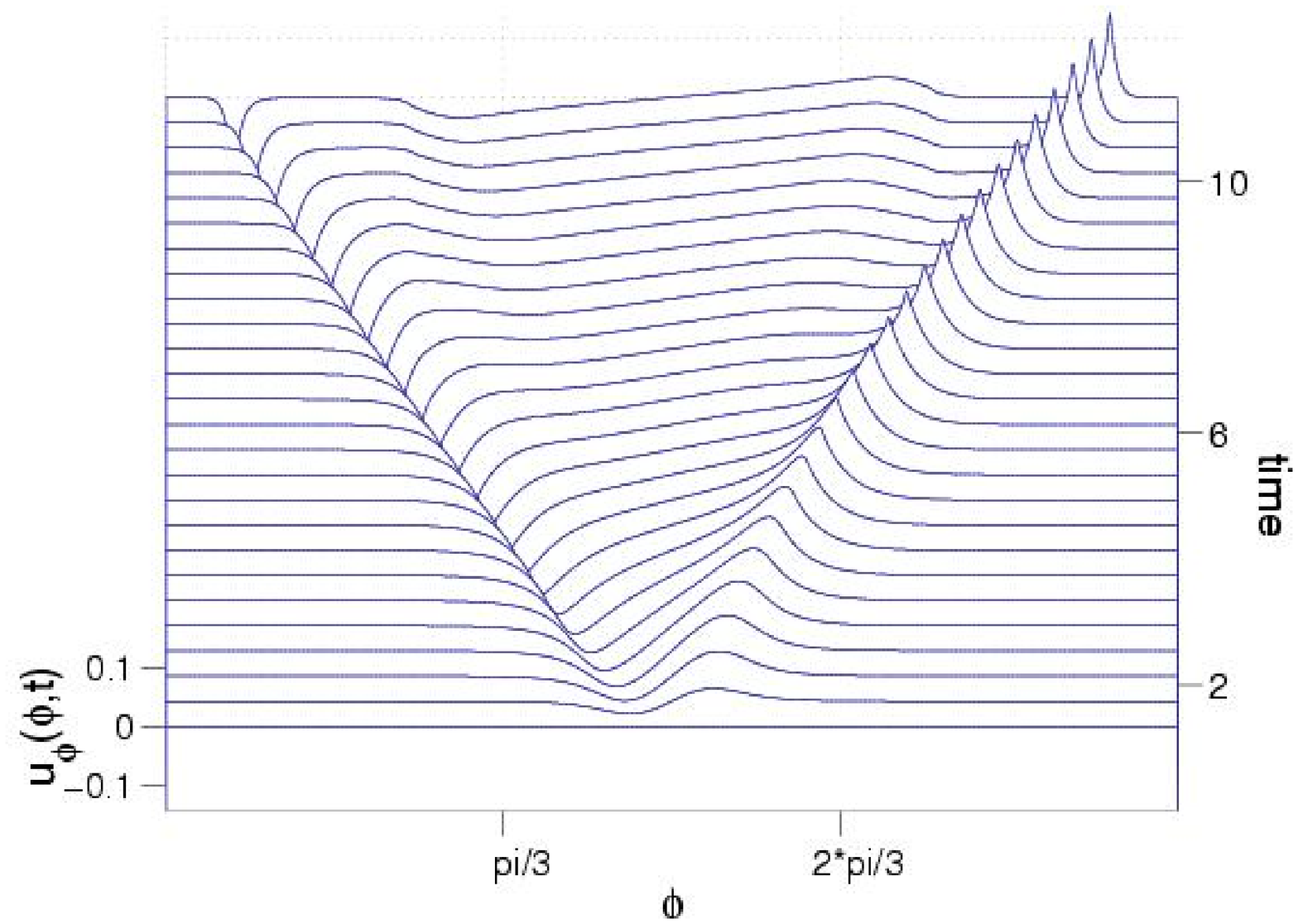}
\includegraphics[scale=0.4]{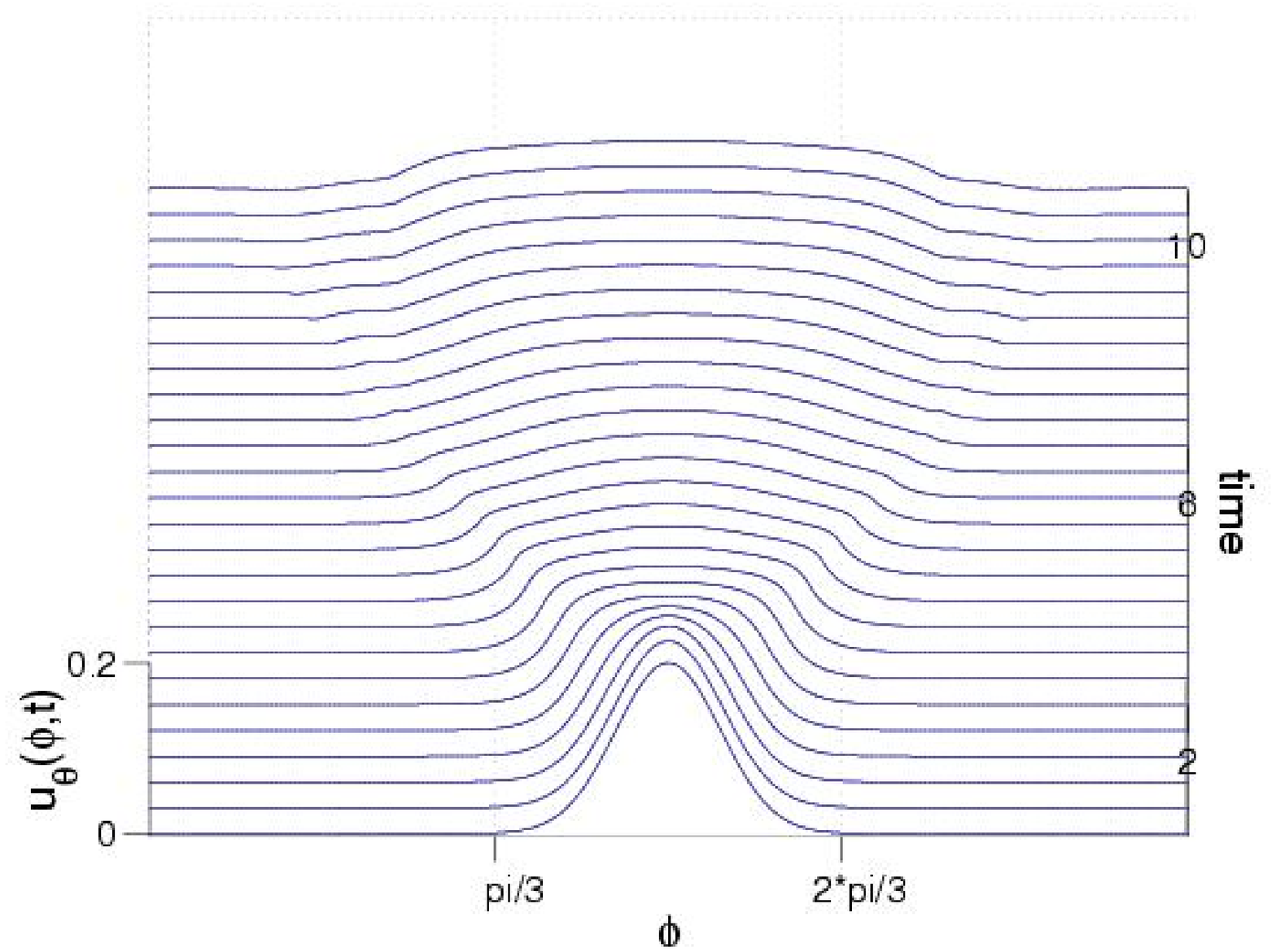}
\caption{The evolution when the initial meridional velocity
$u_\phi$ is zero and the initial azimuthal velocity
$u_\theta$ is a Gaussian.  The larger picture shows the
magnitude of the velocity.  Two rotating Puckons have emerged
at the time shown, and two more are in the process of emerging
from the Gaussian.}
\label{rotIVP-fig}
\end{figure}

\paragraph{Rotating Puckons.}
Now we consider rotating Puckons ($u_\theta\neq 0$).  Figure \ref{rotIVP-fig}
shows the PDE evolution when the initial meridional velocity
$u_\phi$ is zero and the initial azimuthal velocity
$u_\theta$ is a Gaussian.
Two rotating Puckons have emerged from the Gaussian by the
time shown, and they are moving toward
opposite poles.  The Puckons each have small but nonzero
azimuthal velocities.\par

\begin{figure}
\centering
\includegraphics[scale=0.4]{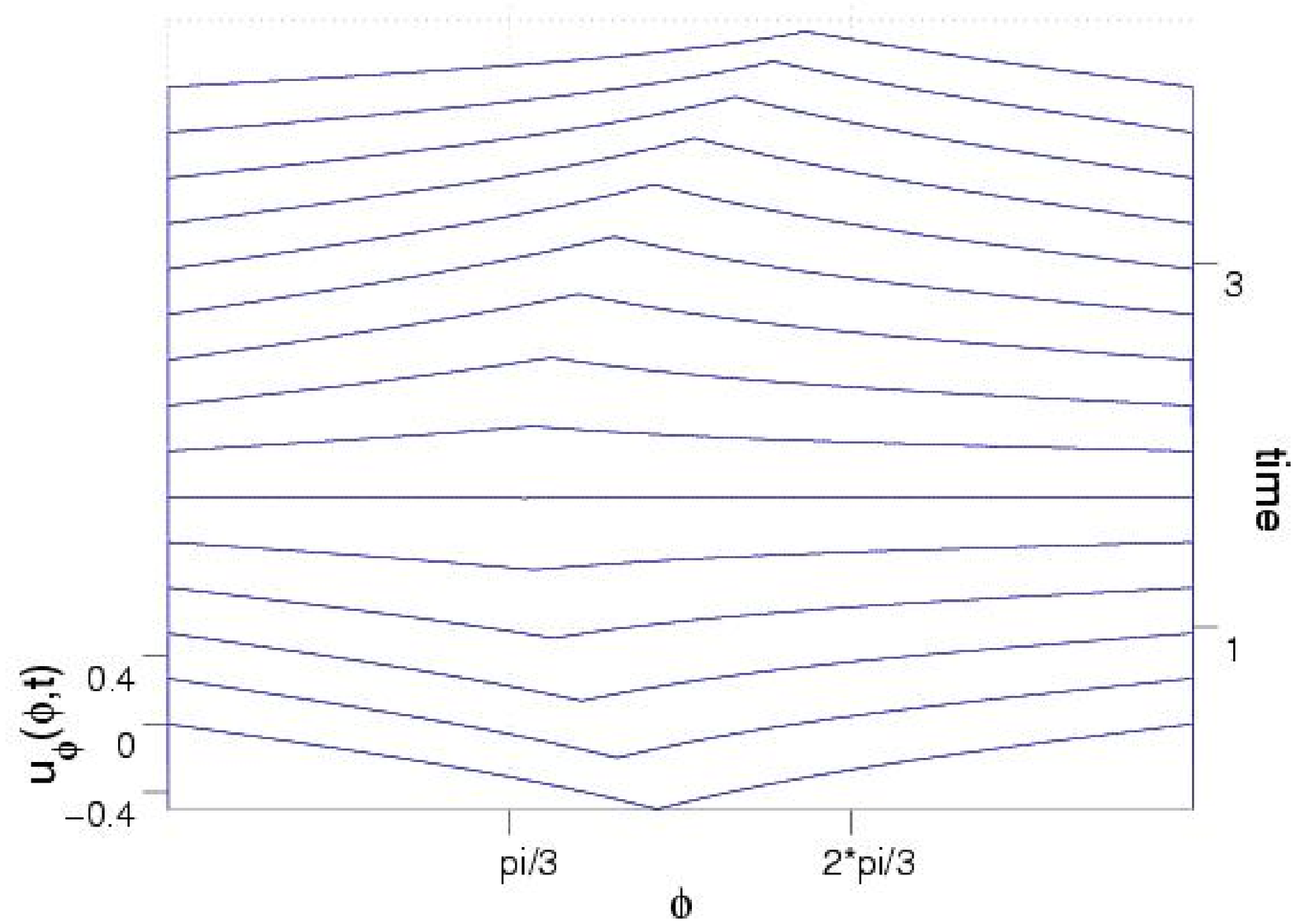}
\includegraphics[scale=0.4]{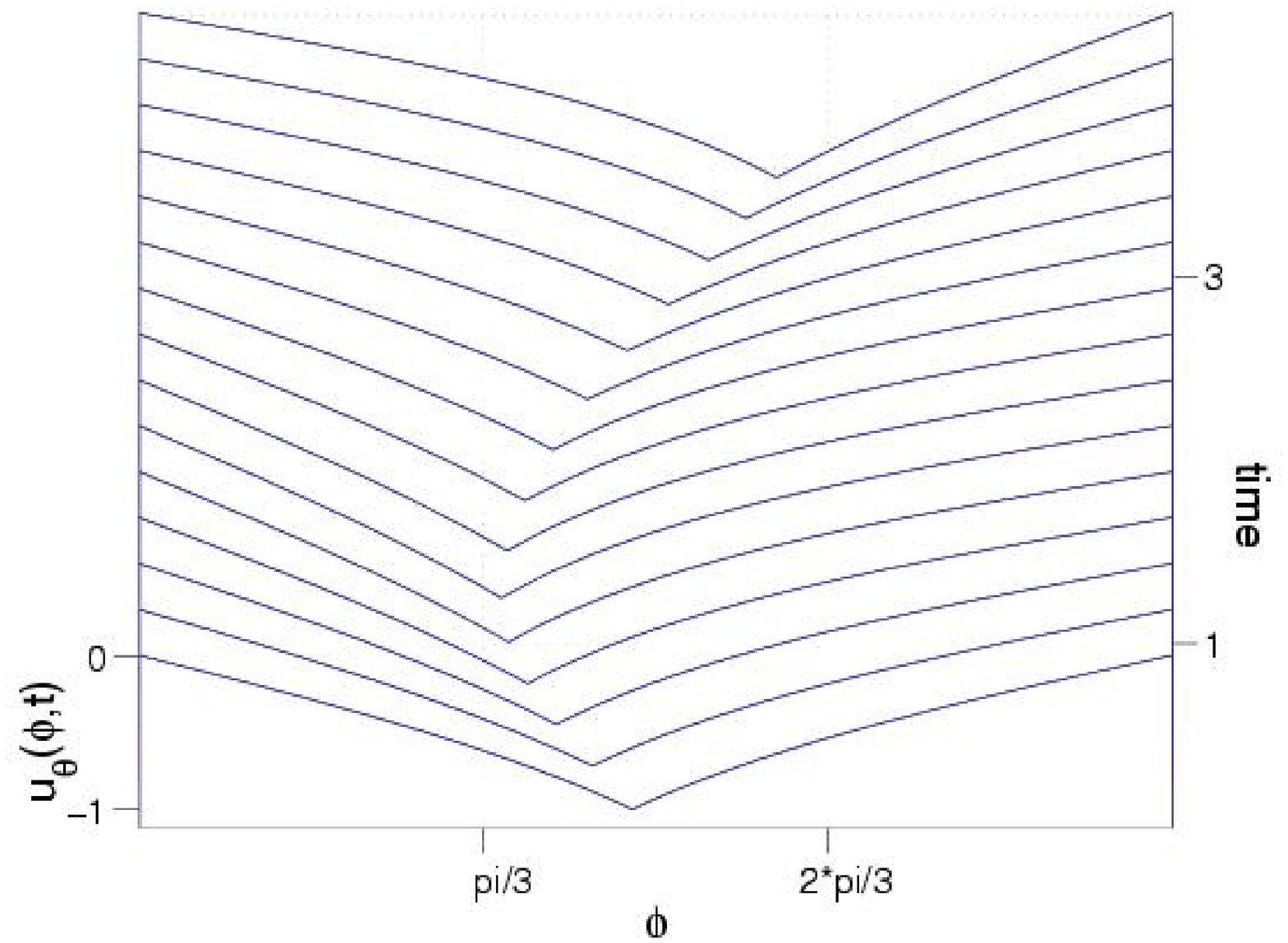}

\includegraphics[scale=0.4]{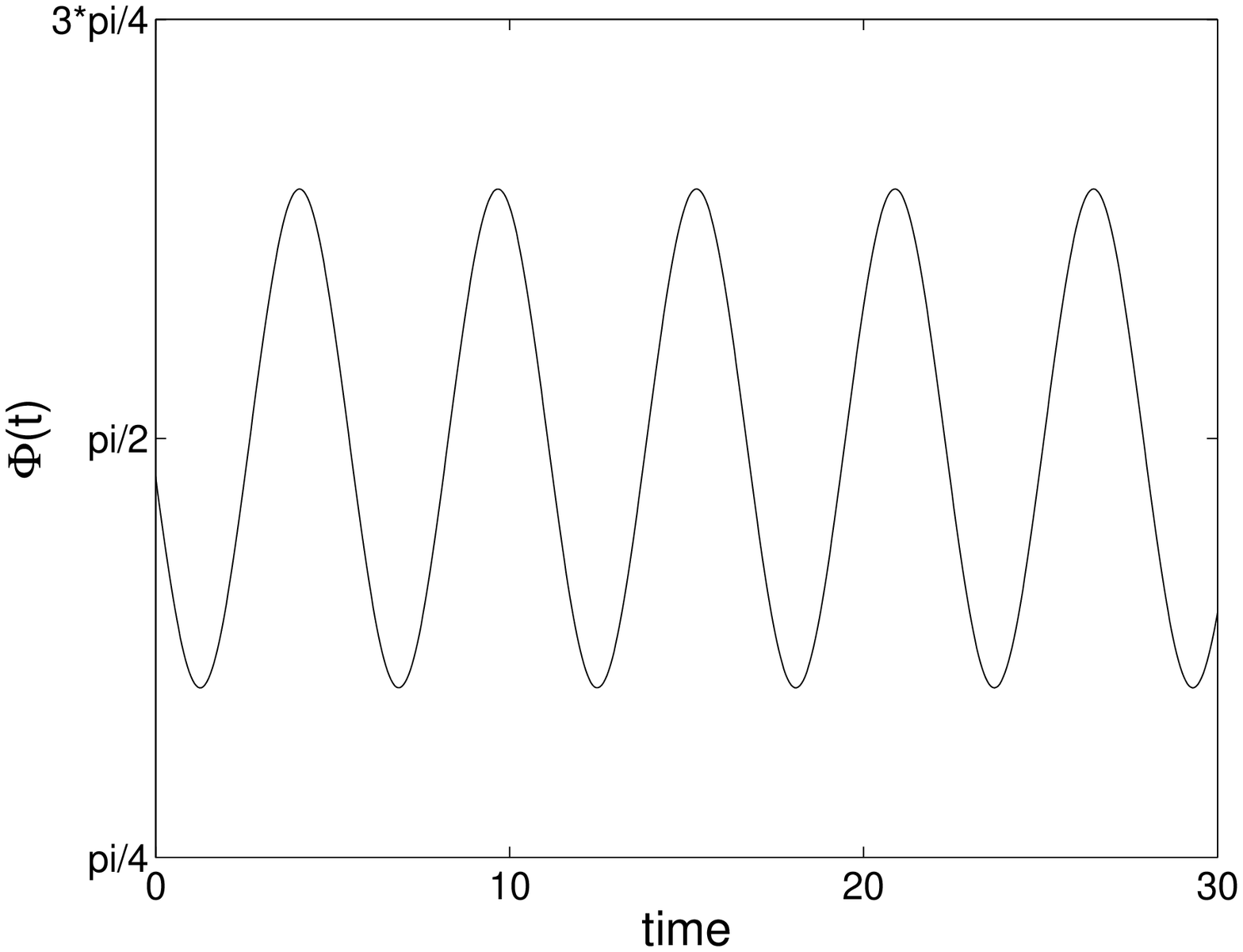}
\includegraphics[scale=0.4]{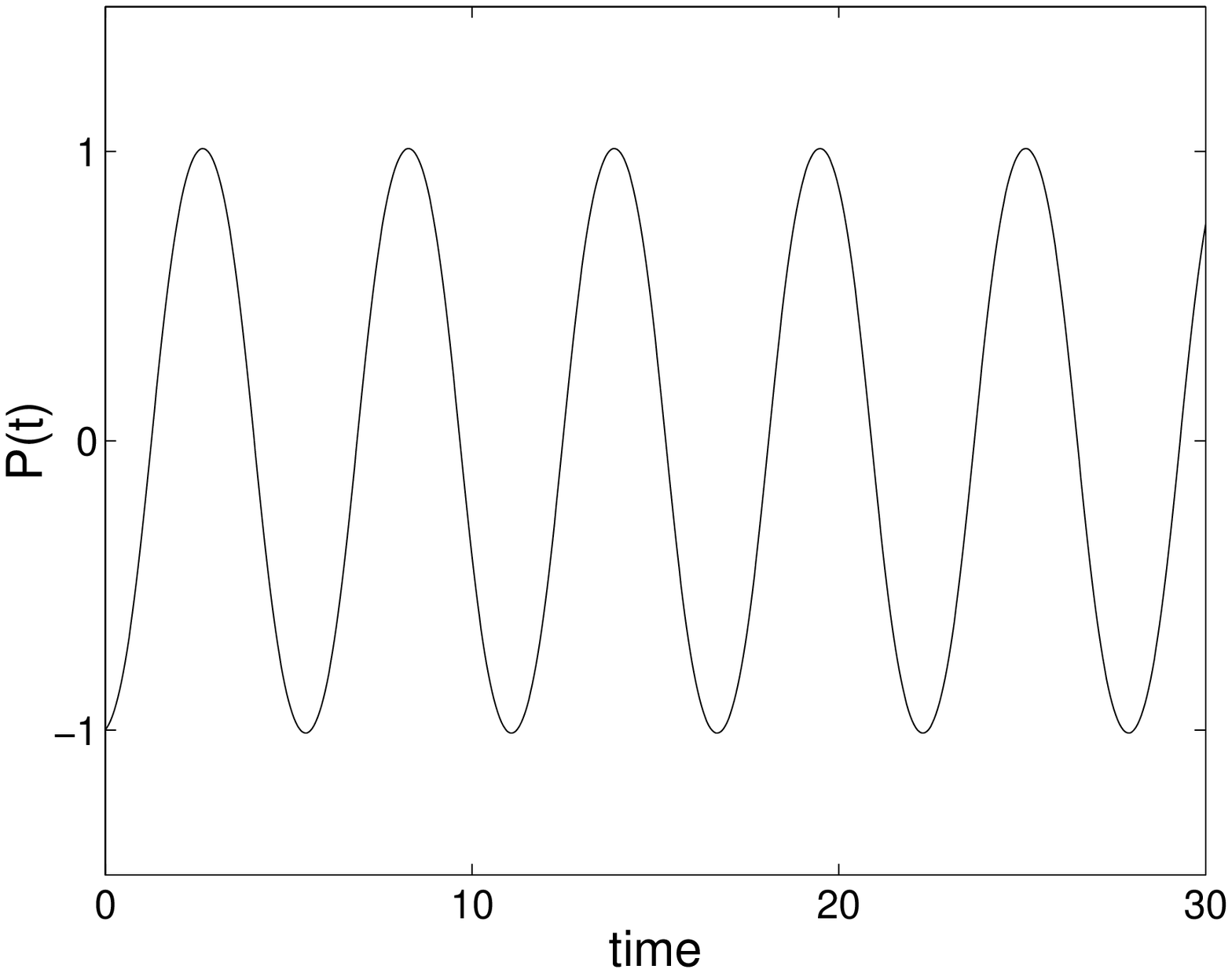}
\caption{A single rotating Puckon for length scale $\alpha=1$.  
  The initial parameters are $\Phi=1.5$, $P=-1$, $M=-2$.
  The PDE simulation
  shows about half of a period of the Puckon's motion, and
  the ODE simulation shows about five periods of the 
  Puckon's motion.  The meridional and azimuthal velocity
  components $u_\phi$ and $u_\theta$ are shown for the
  PDE simulation.}
\label{onerot-fig}
\end{figure}

Finally, Figure \ref{onerot-fig} shows an ODE simulation of a single rotating
Puckon when the length scale is $\alpha=1$ and the canonical 
variables are initially $\Phi=1.5$, $P=-1$, $M=-2$.
As discussed in
Subsection \ref{BasicRot-subsec}, the basic rotating Puckon
moves between minimum and maximum colatitudes.
The Puckon is initially moving toward the North Pole,
and one can see that its meridional velocity vanishes
as it changes its direction and moves toward the South Pole.  
Its azimuthal velocity is negative throughout, i.e., it does
not change its sense of rotation.  Numerically, the Puckon
moves between minimum and maximum colatitudes of (to five
significant digits) 1.1033 and 2.0384, in comparison with
the corresponding values of  1.1031 and 2.0385 predicted by the formula given in
Subsection \ref{BasicRot-subsec}.  Also, the numerical result for the period of
the Puckon's motion (averaged over the first
five periods) agrees to six significant digits (5.60858) with
the value given in Subsection \ref{BasicRot-subsec}. 
Thus, these numerical results for the ODE dynamics are accurate to between four
and six significant figures.

\newpage

\section{Generalising to Other Surfaces}\label{OtherSurfaces-sec}
We now wish to extend these Hamiltonian reductions for diffeons on the
sphere to other surfaces. For the moment, we retain
the rotational symmetry.
\subsection{Rotationally Symmetric Surfaces}\label{rotsym}
Any surface $\Sigma$ with an isometric action of $\sphere{1}$ has
rotationally invariant coordinate charts on which the metric is
given by
\[
g=\d r^2+\psi(r)^2\d\theta^2
\]
where $(r,\theta)\in(r_{min},r_{max})\times(0,2\pi)$. We may take
$r_{min}=0$ to describe a fixed point of the motion. Let $\grad$ be
the Levi-Civita connection with respect to this metric and set
\[
\mathbb{I}=(\id+\Delta^\grad).
\]
By the Bochner-Wietzenb\"ock theorem on 1-forms we compute
\begin{eqnarray*}
\mathbb{I}&=&\id+\Delta^d-\mathrm{Ric}\\
          &=&\left(\frac{\psi+\psi''}{\psi}\right)\id+\Delta^\d.
\end{eqnarray*}
We also have the explicit relation,
\[
\Delta^\d(a(r)\d
r+b(r)\d\theta)=-\Diff{r}\left(\frac{1}{\psi}\Diff{r}(\psi
a(r))\right)\dr -\psi\Diff{r}\left(\frac{1}{\psi}\Diff{r}(
b(r))\right)\d\theta.\]
 Suppose that $\bu$ is a
rotationally invariant vector field on $\Sigma$. That is, suppose
\[
\bu=u^r(r)\partial_r+u^\theta(r)\partial_\theta
\,,
\]
then the singular momentum solution ansatz for EPDiff, 
\begin{equation}\label{Moment0}
\mathbb{I}\bu_\flat=\frac{\delta(r-R)}{\psi(R)}\d
r+\psi(R)\delta(r-R)\d\theta
\,,\end{equation}
is a system of ordinary differential equations in $r$. We may
solve this system to obtain two rotationally invariant, continuous
functions $G^r,G^\theta$ of $r$ and $R$, within our coordinate
chart, which are symmetric in the two variables and which satisfy
\[
\mathbb{I}\left(G^r(r,R)\d r+G^\theta(r,R)\d
\theta\right)=\frac{\delta(r-R)}{\psi(R)}\d
r+\psi(R)\delta(r-R)\d\theta
\,.
\]
The two Green's functions $G^r,G^\theta$ are related by
\[
G^\theta(r,R)=\psi(r)\psi(R)G^r(r,R).
\]
They also have a jump in the derivative along the diagonal $r=R$.
Thus the solution of (\ref{Moment0}) for the velocity  $\bu$ is
\begin{eqnarray*}
\bu&=&\left(G^r(r,R)\d
r+\psi(r)\psi(R)G^r(r,R)\d\theta\right)^\sharp\\
   &=&{G^r(r,R)}\Diff{r}+\psi(R)\frac{G^r(r,R)}{\psi(r)}\Diff{\theta}
\,.
\end{eqnarray*}
The EPDiff equations on the rotationally symmetric surface $\Sigma$ are
\begin{eqnarray}
\label{SigCH1} \bm&=&\mathbb{I}\bu_\flat=\sum_{i=1}^N\frac{\delta(r-R_i)}{\psi(R_i)}\left(P_i\d r+M_i\d\theta\right)\\
\label{SigCH2} 0&=&\diff{\bm}{t}+\mathrm{ad}_{\bu}^*\bm\,,
\end{eqnarray}
and these equations extremise
\begin{eqnarray*}
\ell[\bu]&=&\frac{1}{2}\int_\Sigma (|\bu|_g^2+|\grad\bu|_g^2)\d vol_g\\
         &=&\frac{1}{2}\int_\Sigma \inner{\bu}{\mathbb{I}\bu}\d vol_g\\
         &=&\frac{1}{2}\int_\Sigma \bm(\bu)\d vol_g
\,.
\end{eqnarray*}
After the Legendre transformation we arrive at the Hamiltonian
\begin{eqnarray*}
H[\bm]&=&\int_\Sigma \bm(\bu)\d vol-\ell[\bu]\\
      &=&\frac{1}{2}\int_\Sigma \bm(\bu)\d vol.
\end{eqnarray*}
 We already know that the solution to (\ref{SigCH1}) is
\[
\bu_\flat=\sum_{i=1}^N\frac{G^r(r,R_i)}{\psi(R_i)}\Big(P_i\psi(R_i)\d
r +{M_i}\psi(r)\d\theta\Big),
\]
that is,
\begin{eqnarray*}
\bu&=&\sum_{i=1}^N\frac{G^r(r,R_i)}{\psi(r)\psi(R_i)}\left(P_i\psi(R_i)\psi(r)\Diff{r}
+{M_i}\Diff{\theta}\right)\\
   &=:&u_r\Diff{r}+u_\theta\Diff{\theta}
\,.
\end{eqnarray*}
Because $\bm$ is only supported within our coordinate
chart and is invariant under the action of $\sphere{1}$, we have
\[
H(\bm)=\frac{1}{2}2\pi\sum_{i=1}^N\left(P_iu_r(R_i)+M_iu_\theta(R_i)\right)
\]
where $2\pi$ arises as the volume of $\sphere{1}$.
 Choose a vector
field $w=f(r,\theta)\Diff{r}+g(r,\theta)\Diff{\theta}$. Again
since $\bm$ is supported within our chart, we know that
\[
\int_{\sigma}\bm(w)\d
vol_g=\sum_i\int_0^{2\pi}\left(P_if(R_i,\theta)+M_ig(R_i,\theta)\right)\d\theta.
\]
We also have
\[
\left[w,\bu\right]=\left(fu_r'-u_r\partial_rf-u_\theta\partial_\theta
f\right)\Diff{r}+\left(fu_\theta'-u_r\partial_rg-u_\theta\partial_\theta
g\right)\Diff{\theta}.
\]
Consequently, the EPDiff equations yield
\begin{eqnarray*}
0&=&\int_\Sigma \left(\Diff{t}(\bm(w))+\bm([w,\bu])\right)\d vol_g\\
 &=&\int_0^{2\pi}\Diff{t}\left(P_if(R_i,\theta)+M_ig(R_i,\theta)\right)\d\theta\\
 &+&\int_0^{2\pi}\left(P_i\left(fu_r'-u_r\partial_rf-u_\theta\partial_\theta
f\right)+M_i\left(fu_\theta'-u_r\partial_rg-u_\theta\partial_\theta
g\right)\right)\big\vert_{r=R_i}\d\theta\\
 &=&\int_0^{2\pi}\left(\dot{P}_if(R_i,\theta)+P_i\dot{R}_i\partial_rf(R_i,\theta)+\dot{M}_ig(R_i,\theta)+M_i\dot{R_i}\partial_rg(R_i,\theta)\right)\d\theta\\
 &+&\int_0^{2\pi}\left(P_i\left(fu_r'-u_r\partial_rf\right)+M_i\left(fu_\theta'-u_r\partial_rg\right)\right)\big\vert_{r=R_i}\d\theta.
\end{eqnarray*}
Comparing coefficients of $f,\partial_rf,g\partial_rg$ now yields
\begin{eqnarray*}
0&=&\dot{P}_i+P_iu_r'(R_i)+M_iu_\theta'(R_i)\\
0&=&P_i\dot{R}_i-P_iu_r(R_i)\\
0&=&\dot{M}_i\\
0&=&M_i\dot{R}_i-M_iu_r(R_i)
\end{eqnarray*}
Now since
\[
u_r(R_i)=\sum_{j=1}^NG^r(R_i,R_j)
\]
and
\[
u_r'(R_i)=\sum_{j=1}^N\diff{G^r}{r}(R_i,R_j)
\]
 we see that
\begin{eqnarray*}
\Diff{R_i}\sum_{j=1}^N
u_r(R_j)&=&\Diff{R_i}\sum_{j,k=1}^NG^r(R_j,R_k)\\
&=&\sum_{j,k=1}^N\left(\diff{R_j}{R_i}\left.\diff{G^r(r,R)}{r}\right\vert_{r=R_j,R=R_k}
+\diff{R_k}{R_i}\left.\diff{G^r(r,R)}{R}\right\vert_{r=R_j,R=R_k}\right)\\
&=&\sum_{j,k=1}^N\left(\diff{R_j}{R_i}\left.\diff{G^r(r,R)}{r}\right\vert_{r=R_j,R=R_k}
+\diff{R_k}{R_i}\left.\diff{G^r(r,R)}{r}\right\vert_{r=R_k,R=R_j}\right)\\
& & \hbox{since }G^r\hbox{ is symmetric in its arguments}\\
&=&\sum_{j,k=1}^N\left(\delta_{ij}\left.\diff{G^r(r,R)}{r}\right\vert_{r=R_j,R=R_k}
+\delta_{ki}\left.\diff{G^r(r,R)}{r}\right\vert_{r=R_k,R=R_j}\right)\\
&=&\sum_{k=1}^N\left.\diff{G^r(r,R)}{r}\right\vert_{r=R_i,R=R_k}
+\sum_{j=1}^N\left.\diff{G^r(r,R)}{r}\right\vert_{r=R_i,R=R_j}\\
&=&2\sum_{j=1}^N\left.\diff{G^r(r,R)}{r}\right\vert_{r=R_i,R=R_j}\\
&=&2u_r'(R_i).
\end{eqnarray*}
Similarly
\[
u_\theta'(R_i)=\frac{1}{2}\Diff{R_i}\sum_{j=1}^N u_\theta(R_j).
\]
 This finally yields
\begin{eqnarray}
\label{Sig1}\dot{P}_i&=&-\frac{1}{2}\Diff{R_i}\sum_{j=1}^N\left(P_ju_r(R_j)+M_ju_\theta(R_j)\right)\\
\label{Sig2}\dot{R}_i&=&u_r(R_i)\\
\label{Sig3}\dot{M}_i&=&0.
\end{eqnarray}
 Thus, we have proved the following realisation of the momentum map in
\cite{HoMa2004}.
\begin{prop1} [Canonical equations for rotationally symmetric diffeons]  The
parameters $P_i,R_i$ for diffeons on a rotationally symmetric surface $\Sigma$
with constant positive curvature and metric given by $g=\d
r^2+\psi(r)^2\d\theta^2$ satisfy Hamilton's canonical equations with
Hamiltonian given by
\begin{eqnarray*}
H(\mathbf{P},\mathbf{R};\mathbf{M})
&=&\frac{1}{2\pi}\left(\int_\Sigma\bm(\bu)\d vol-\ell[\bu]\right)\\
&=&\frac{1}{2}\sum_{i=1}^N\left(P_iu_r(R_i)+M_iu_\theta(R_i)\right)\\
&=&\frac{1}{2}\sum_{i,j=1}^N\frac{G^r(R_i,R_j)}{\psi(R_i)\psi(R_j)}
\left(P_iP_j\psi(R_i)\psi(R_j)+M_iM_j\right).
\end{eqnarray*}
A solution to (\ref{Sig1},\ref{Sig2},\ref{Sig3}) is an example of
an $N$-diffeon on a rotationally symmetric surface $\Sigma$.
\end{prop1}\par
We notice that the critical points of $H$ are those
$\mathbf{R}=(R_1,...,R_n)$ such that $G^r(R_i,R_j)=0$ for all
$i,j$. But the value of $H$ at these critical points is always $0$
since $H$ depends on $G^r$.
\begin{rmk1}
The only 1-dimensional Lie Groups are $\reals$ and $\sphere{1}$,
and while we have considered the isometric action of $\sphere{1}$
on a surface, we have to be more careful with translation
invariance because the group ceases to be compact. However we can
consider $\theta$, which parameterises each orbit $r=const$,
taking values in $(-L,L)$ rather than $(-\infty,\infty)$ for the
full translation group. This makes the integration finite.
\end{rmk1}
\subsection{Rotationally Invariant Diffeons on Hyperbolic Space}
Hyperbolic space has a richer structure than either the plane, or
the sphere. Indeed, the isometry group of the sphere is $\SO{3}$;
so all isometries are rotations. We have already examined the case
of rotationally invariant diffeons. (These are the Puckons.) The isometry
group of the plane $\reals^2$ is $\sphere{1}\ltimes\reals^2$, the
semi-direct product of rotations and translations. Translational
invariance yields the direct product of the original 1-dimensional
peakons, whereas rotational invariance yields the rotating
circular peakons developed in \cite{HoPuSt2004}. However, the isometry
group of hyperbolic space is $\proj\SL{2}{\reals}$, which is not
compact and contains three different types of isometry. These are the
rotational, translational and horolational subgroups.

\paragraph{The Hamiltonian for hyperbolic
$N$-diffeons with rotational symmetry.} 
We have already done most of the work for the rotationally symmetric case.
All that remains is to write the hyperbolic metric in a conformally flat way and
from it deduce the Green's functions $G^r$ and $G^\theta$. The model is
familiar: we use the Poincar\'e disc $D=\{(x,y)\in\reals^2|\ x^2+y^2<1\}$ in
polar coordinates with the metric
\[
g_\hyperbolic=\frac{4}{(1-r^2)^2}\left(\dr^2+r^2\d\theta^2\right),
\]
i.e. the conformal parameter is \[ \rho(r)=\frac{2}{1-r^2}\,.\] The
Bochner-Wietzenb\"ock formula states that on $\Tang{D}$
\[
\Delta^\d=\Delta^\grad-\id\,,
\]
so that
\begin{equation}\label{hypweitz}
\mathbb{I}=\id+\Delta^\grad=2\id+\Delta^\d.
\end{equation}
We wish to find Green's functions $G^r$ and $G^\theta$ such that
\[
\mathbb{I}(G^r(r,R)\dr+G^\theta(r,R)\d\theta)=\delta(r-R)\left(\frac{1}{R}\dr+R\d\theta\right).
\]
We also ask that these Green's functions be finite at the origin, continuous
and symmetric in their variables. Explicitly, we have
\[
\Delta^\d\left(a(r)\dr+b(r)\d\theta\right)=-\Diff{r}\left(\frac{1}{\rho^2r}
\Diff{r}(ra(r))\right)\dr-r\Diff{r}\left(\frac{1}{\rho^2r}\diff{b}{r}\right)
\d\theta.
\]
Upon setting
\[
F(r)=2-2r^4+8\log(r)r^2+r^2
\]
we write
\[
G^r(r,R)=\frac{\rho(r)^2\rho(R)^2}{16}\frac{\min(r,R)}{\min(r,R)}F(\max(r,R)),
\]
and hence
\[
G^\theta(r,R)=RrG^r(r,R).
\]
Thus, we have shown the following.

\begin{prop1}[Hamiltonian for rotationally invariant hyperbolic $N$-diffeons]
\hfill\\The Hamiltonian for rotationally invariant
$N$-diffeons on hyperbolic space is given by
\begin{eqnarray*}
H(\mathbf{P},\mathbf{R};\mathbf{M})&=&\frac{1}{2}\sum_{i,j=1}^N\frac{1}{\rho(R_i)^2R_i\rho(R_j)^2R_j}
\left(P_iP_jR_iR_jG^r(R_i,R_j)+\frac{M_iM_j}{R_iR_j}G^\theta(R_i,R_j)\right)\\
&=&\frac{1}{2}\sum_{i,j=1}^N\frac{G^r(R_i,R_j)}{\rho(R_i)^2R_i\rho(R_j)^2R_j}
\left(P_iP_jR_iR_j+M_iM_j\right).
\end{eqnarray*}
\end{prop1}

\paragraph{The hyperbolic diffeon.}
 Let us examine the behaviour of the basic rotationally symmetric hyperbolic
diffeon. The Hamiltonian is
\begin{eqnarray*}
H(P,R)
&=&\frac{1}{2}\frac{G^r(R,R)}{\rho(R)^4R^2}
\left(P^2R^2+{M^2}\right)\\
&=&\frac{F(R)}{32R^2}
\left(P^2R^2+{M^2}\right).
\end{eqnarray*}
The function $F(R)/R^2$ is positive and continuous on $(0,1)$, and it
has a singularity at $R=0$. Its derivative is negative on $(0,1)$
and its image is $(1,\infty)$. Consequently, 
$F(R)/R^2$ is an invertible function $(0,1)\mapping (1,\infty)$.
\par
Assuming $M\neq0$, on the level set $H(P,R)=K^2$ 
we find that
\[
P^2=\frac{32K^2}{F(R)}-\frac{M^2}{R^2}\,.
\]
So $\dot{R}=0$, whenever
\[
\frac{F(R)}{R^2}=\frac{32K^2}{M^2}.
\]
Hence, whenever $32K^2>M^2>0$, precisely one
turning point exists for $R$. This proves the following.

\begin{prop1}
Rotating diffeons with $M^2>0$ on a rotationally symmetric hyperbolic surface
do not exhibit periodic behaviour. 
\end{prop1}

\par For the irrotational diffeon, $M=0$ and we have
\[
P^2=\frac{32K^2}{F(R)}\,.
\]
Since $F(R)>0$ for all $R\in [0,1]$, the irrotational diffeon has no turning
points of $R$ at all.
\subsection{Horolationally Invariant Diffeons on Hyperbolic
Space} We now turn to a subtly different problem. So far we have
been using solely the rotation group (the circle) to produce
symmetric diffeons. Now we consider a different subgroup of
hyperbolic isometries. This time it is expedient to use the
upper-half-plane model of hyperbolic geometry. That is,
\[
\hyperbolic=\{(x,y)\in\reals^2|y>0\}
\]
with the metric
\[
g_\hyperbolic=\frac{\d x^2+\d y^2}{y^2}\,.
\]
The group we consider is the horolation group $(x,y)\mapsto
(x+a,y)$. This is a unique type of isometry in planar geometries
and is, figuratively speaking, a ``rotation about infinity". We
can see immediately that the orbits of this group action are the
lines $y=const$, and that we seek diffeons which are independent
of $x$. In this situation by (\ref{hypweitz}),
\[
\mathbb{I}(a(y)\d x+b(y)\d
y)=\frac{1}{y}\left(2ya-\diff{^2a}{y^2}y^3-2\diff{a}{y}y^2\right)\d
x+\left(2b-\diff{^2b}{y^2}y^2-2\diff{b}{y}y\right)\d y
\]
Thus the solution to
\[
\mathbb{I}(G(y,Y)\d x)=\delta(y-Y)\d x
\]
is
\[
G(y,Y)=\frac{\min(y,Y)}{3\max(y,Y)^2},
\]
and this solution also solves
\[
\mathbb{I}(G(y,Y)\d y)=\delta(y-Y)\d y.
\]
Next, we consider the compactness issue. Since the coordinate $x$ ranges
from $-\infty$ to $\infty$ we know that functions that only
involve $y$ cannot be integrated across all of $\hyperbolic$. Thus
the Diffeon Hamiltonian cannot exist over the whole space and the
theory of section \ref{rotsym} does not apply. Let us restrict
ourselves to the strip $B_L=\{(x,y)|y>0,|x|<
L\}\subset\mathcal{H}$. By doing this we can apply similar
calculations to those used in section \ref{rotsym} to $B_L$, but we
must apply them only in the case of vector fields which are
tangent to the boundary $x=\pm L$.
\par
Thus if
\[
\bm=\sum_{i=1}^N\delta(y-Y_i)Y_i^2(M_i\d x+P_i\d y)
\]
then we arrive at the Hamiltonian
\begin{eqnarray*}
H(\mathbf{P},\mathbf{Y})&=&L\sum_{i,j=1}^N
G(Y_i,Y_j)Y_i^2Y_j^2(P_iP_j+M_iM_j)\\
&=&L\sum_{i,j=1}^N
\frac{\min(Y_i,Y_j)}{3\max(Y_i,Y_j)^2}Y_i^2Y_j^2(P_iP_j+M_iM_j)\\
&=&L\sum_{i,j=1}^N
\frac{\min(Y_i,Y_j)^3}{3}(P_iP_j+M_iM_j)\\
\end{eqnarray*}
and the condition that $\dot{M}_i=0$.\par For the 1-diffeon, the
Hamiltonian is
\[
H(P,Y)=L\frac{Y^3}{3}(P^2+M^2)
\]
so for $H(P,Y)=const=K$, the only turning point of $Y$ is at
\[
Y=\sqrt[3]{\frac{3K}{LM^2}}
\]
or at 0, if $M=0$.
\begin{rmk1}
Joining the edges of $B_L$ together yields a surface called
Gabriel's Horn, whereby the translation invariant diffeons on
$B_L$ become rotationally invariant diffeons on Gabriel's
horn.
\end{rmk1}
\subsection{Translation Invariant Diffeons on Hyperbolic Space}
Translations on the Hyperbolic plane are characterised by having
two ideal fixed points. The version of the metric we use in this case for the
upper half plane is
\[
g_\hyperbolic=\d r+\cosh^2r\d\theta^2\,,
\]
where the subsets formed by $r=const$ are the orbits of the
translation. Again, the orbits are not compact, so we limit
ourselves to the strip
$B_L=\{(r,\theta)\in\reals^2|r>0,|\theta|<L\}\subset\mathcal{H}$.
\par
In this situation, the Green's function $G(r,R)$ solving
\[
\mathbb{I}G(r,R)\d r=\frac{\delta(r-R)}{\cosh(R)}\d r
\]
is given by
\[
G(r,R)=\frac{1}{2}\tanh(\min(r,R))\cosh(\max(r,R))+\cosh(r)\cosh(R)\arctan(e^{\min(r,R)}).
\]
Due to the complicated form of the Green's function, we will not
write down the explicit form of the Hamiltonian for the 
$N$-diffeon. However, the Hamiltonian for the 1-diffeon is,
\begin{eqnarray*}
H(P,R;M)&=&\frac{1}{2}\frac{G(R,R)}{\cosh^2R}
\left(P^2\cosh^2R+M^2\right)\\
&=&\frac{1}{4}\left(\tanh
R\,\mathrm{sech}\,R+2\arctan(e^{R})\right)
\left(P^2\cosh^2R+M^2\right).\\
\end{eqnarray*}
In this case
\begin{eqnarray*}
\diff{H}{P}&=&\frac{1}{2}\left(\tanh
R\,\mathrm{sech}\,R+2\arctan(e^{R})\right)P\cosh^2R\,,\\
\diff{H}{R}&=&\frac{1}{2}\left(P^2\left(\cosh{R}+\sinh{2R}
\arctan(e^{R})\right)+M^2\,\mathrm{sech}^3\,R\right)
\,.
\end{eqnarray*}
 It can be shown that there are no equilibria for the
motion unless $M=0$, where the line $P=0$ is the critical point set
for $H$. However,
\[
H(0,R)=0\,,
\]
for each value of $R$. So the line $P$ consists entirely of
stationary points, and since away from this line no
points exist for which $\dot R=0$, there can be no periodic behaviour of
1-diffeons in this case.

\rem{
This raises the question of the r\^ole that the
curvature plays in the behaviour of diffeons. For surfaces of
constant positive curvature,i.e the sphere, all diffeons manifest
periodic behaviour. In the case of the plane, periodic behaviour
is admitted, but not all diffeons on the plane need be periodic.
Finally, as we have seen for the hyperbolic plane, no periodic
behaviour is admitted.
}
\paragraph{Extending EPDiff on Hyperbolic Space}
Our study of EPDiff on hyperbolic spaces so far shows that the behaviour of
hyperbolic diffeons seems less interesting than the rich dynamical 
structure of multiple Puckons interacting on the sphere. This is because
hyperbolic diffeons will bounce only once before heading off to infinity,
in the cases we have studied so far.\par
However, hyperbolic space is related intimately with the study of curves of
genus at least 2: it forms the universal cover for all such Riemann
surfaces. So far, we have been unable to examine the case of curves of genus
$\ge 2$ simply because their isometry groups are so small (usually
discrete) that we simply cannot find a 1-parameter subgroup by which we may
reduce the EPDiff equations to ODEs.\par 
Any high genus Riemann surface can be realised as the quotient of the
hyperbolic plane by a tiling, or hyperbolic lattice \cite{Wo1990}. The
complex structure of the Riemann surface depends upon the shape of the
lattice. A surface
$\Sigma$ of genus
$\gamma\ge 2$ is determined topologically by the symmetry group of the
tiling; but there are a variety of complex structures which may be put upon
$\Sigma$ to turn it into a Riemann Surface. If $\Gamma$ is the symmetry
group of the tiling, then the space of complex structures
which may be put on $\Sigma$ is given by the space
\[
T_\Gamma=\frac{\Gamma\hbox{-invariant quasiconformal maps of }
\mathbb{D}}{\hbox{Hyperbolic isometries of } \mathbb{D}}.
\]
Here $\mathbb{D}$ is the Poincar\'e disc (the unit disc in the complex
plane).  In the space $T_\Gamma$ a quasiconformal map is a generalisation of
a conformal map in which although angles are not preserved, the angular
dilation is uniformly bounded, see \cite{Ahl,GaHa2001}.\par
The inclusion of a subgroup $\Gamma'$ in $\Gamma$ induces a natural
inclusion of $T_\Gamma$ in $T_{\Gamma'}$, so it makes sense to consider
  the universal Teichm\"uller space as the Teichm\"uller space $T_1$
  associated with the trivial group. The universal Teichm\"uller space is endowed with the Weil-Petersson metric, thus
providing the framework essential for the theory of
EPDiff. Furthermore, the restriction of the maps given in the definition
of Teichm\"uller space to the boundary of the disc provides us with
the essential method by which we reduce the EPDiff PDEs to ODEs. This
study is beyond the scope of this paper and we present the results in
the next, \cite{HMS}.
\section{EPDiff on Higher Dimensional Manifolds with
Symmetry}\label{warp-sec} 
Our calculation for surfaces in the previous sections impose a
rotational symmetry of the solutions. That is, the solutions are invariant
under a circle action, and this effectively reduces $\mathbb{I}$ in
(\ref{hypweitz}) to an ordinary differential operator. As a result, we are
able to find $G^r$ and $G^\theta$ which are invariant under the group
action, continuous and have a jump in the derivative along an orbit. This
motivates the study of higher dimensional Riemannian manifolds
$(\Sigma,g)$ which possess an isometric action of the Lie group $G$ such
that the quotient space $\Sigma/G$ is a 1-dimensional orbifold.
\subsection{Principal Bundles over 1-Dimensional Manifolds} We
consider first the case of principal bundles over 1-dimensional
manifolds. It is a well known fact that 1-manifolds are
diffeomorphic to $\reals$, $[0,\infty)$, $[0,1]$ or $\sphere{1}$,
see for example pp55-57 of \cite{MI}. 
Intervals of $\reals$ are all contractible, we know that any
principal bundle over them is trivial. The principal $K$-bundles
of $\sphere{1}$ are up to isomorphism in 1-1 correspondence with
$\pi_1(BK)$, the fundamental group of the classifying space $BK$
of the Lie group $K$. However, by the long exact sequence of
homotopy groups (see \cite{LM}), we know that
$\pi_1(BK)\cong\pi_0(K)$, thus if $K$ is connected, then any
Principal $K$-bundle over $\sphere{1}$ is trivial.\par The upshot
here is that if $\Sigma$ is a Riemannian manifold with a free
isometric action of $K$ and $\Sigma/K$ is a 1-dimensional manifold
$N$, then $\Sigma\cong K\times N$.
\subsection{Warped Products}
Let $K$ be a compact semi-simple Lie group of dimension $n-1$ and
$\Sigma\mapping I\subset\reals$ a principal $K$-bundle over the
open interval $I$ (in fact in what follows, we may also take $I$
to be the circle $\sphere{1}$). Suppose $\Sigma$ has a Riemannian
metric $g$ preserved by $K$, and that coordinates exist such
that the metric has the form of a warped product
\[
g=\d r^2+\psi(r)^2g_K
\]
where $g_K$ is a bi-invariant inner product on $\Tang{K}$ and
$\psi$ is a $K$-invariant function on $\Sigma$, i.e. a function of
$r$ alone. The coordinate $r$ parameterises the orbit of $K$ in
$\Sigma$. Let $\grad$ be the Levi-Civita connection with respect
to $g$.\par

This situation is rich enough to produce interesting results.
In what follows $\{\xi_i\}_{i=1}^n$ will be a $g_K$ orthonormal basis of
$\mathfrak{k}$ and $X_\xi$ the left-invariant vector field
generated by $\xi\in\mathfrak{k}$. We write $X_i=X_{\xi_i}$, these
form an orthonormal frame of $\Tang{K}$; set $\theta^i$ to be the
coframe dual to the $X_i$ . Let $\grad^K$ be the Levi-Civita
connection of $K$ with respect to $g_K$.
\begin{prop1}\label{lielap}
We have
\[
\Delta^{\grad^K}X_{i}=kX_i.
\]
\end{prop1}
This follows from the identity on $K$,
\[
\grad_{X_i}^KX_j=\frac{1}{2}[X_i,X_j]
\,.\]
Thus, $\Delta^{\grad^K}$ will be the Casimir operator
associated with the adjoint representation of $K$ on
$\mathfrak{k}$ applied to the frame $\{X_i\}$. The Casimir however
is just a constant multiple of the identity, the constant $k$
being
\[
k=1-\frac{\dim\mathfrak{z}(K)}{\dim\mathfrak{k}}
\]
where $\mathfrak{z}(K)$ is the Lie algebra of centre of the group
$K$.
\par
The geometry of warped products is reasonably well known, \cite{B,P}. As
an exercise in Riemannian geometry, one may deduce (see pp58-59 of \cite{B})
\begin{prop1}\label{lapform}
Let $f$ be a $K$-invariant function on $\Sigma$ and suppose
$\psi(r)=e^{\lambda(r)}$. The connection Laplacian associated with
$g$ satisfies
\begin{eqnarray*}
\Delta^\grad
f\dr&=&-\left(\diff{^2f}{r^2}+(n-1)\diff{\lambda}{r}\diff{f}{r}-(n-1)\left(\diff{\lambda}{r}\right)^2f\right)\dr\\
\Delta^\grad f
e^\lambda\theta^i&=&-\left(\diff{^2f}{r^2}+(n-1)\diff{\lambda}{r}\diff{f}{r}-\left(\diff{\lambda}{r}\right)^2f-e^{-\lambda}kf\right)e^{\lambda}\theta^i
\end{eqnarray*}
\end{prop1}
These two propositions show us that the $K$-invariant equations
\begin{eqnarray}
\label{I0}\mathbb{I}a^0(r)\dr&=&\frac{\delta(r-R)}{\psi(R)}\dr\\
\label{Ii}\mathbb{I}a^i(r)\psi(r)\theta^i&=&\delta(r-R)\theta^i
\end{eqnarray}
are $n$ ordinary differential equations in $r$. Thus, there are $n$
functions $G^0(r,R),G^i(r,R)$ that solve (\ref{I0}) and (\ref{Ii}),
respectively, are continuous on $r=R$, and are symmetric in the sense that
\begin{eqnarray*}
G^0(r,R)&=&G^0(R,r)\\
{G^i(r,R)}&=&{G^i(R,r)}.
\end{eqnarray*}
Indeed, the second identity in Proposition
\ref{lapform} shows that $G^i$ is {\it independent} of the index $i$. Hence,
instead of $G^i$,we shall write $G^K$.\par
Now let $\bm$ be the measure-valued one-form
\begin{equation}\label{wdiff}
\bm^j=\sum_{i=1}^N\frac{\delta(r-R_i)}{\psi(R_i)}\left(P_i\dr
+M_i\theta^j\right)
\end{equation}
The solution to $\mathbb{I}(\bu_j)_\flat=\bm^j$ is
\begin{eqnarray*}
\bu_j&=&\sum_{i=1}^N \left(P_iG^0(r,R_i)\Diff{r}+\frac{M_i}{\psi(r)\psi(R_i)}G^K(r,R_i)X_j\right)\\
   &=&{u_j^0}(r)\Diff{r}+u_j(r)X_j.
\end{eqnarray*}
Note, we are not using the summation convention here. Also denote
$$w=f^0\Diff{r}+f^\zeta X_\zeta$$ for any $\zeta\in\mathfrak{k}$.
Any vector field on $\Sigma$ will be the sum of such vector fields $w$. Then we
may write the vector field commutation relation, 
\[
[w,\bu_j]=\left(f^0{u_j^0}'-{u_j^0}\partial_rf^0-u_jX_j(f^0)\right)
\Diff{r}-\left({u_j^0}\partial_rf^\zeta+u_jX_j(f^\zeta)\right)X_\zeta+f^0u_j'X_j.
\]
Thus
\begin{eqnarray*}
\int_\Sigma\bm^j(w) \d
vol_P&=&\sum_{i=1}^N\int_K\int_I\frac{\delta(r-R_i)}{\psi(R_i)}\left(P_i
f^0+M_i f^\zeta g_K(X_\zeta,X_j)\right)\psi(r)\dr\d vol_K\\
&=&\sum_{i=1}^N\int_{h\in K}\left(P_i f^0(R_i,h)+M_i
f^\zeta(R_i,h)\zeta^j\right)\d vol_K
\end{eqnarray*}
where $\zeta=\sum_{l=1}^{n-1}\zeta^j\xi_j$. Especially, this yields the
ad$^*$ relation needed for expressing EPDiff in this setting, 
\begin{eqnarray*}
& &\int_\Sigma\bm^j([w,\bu_j])\\
&=&\sum_{i=1}^N\int_{h\in
K}P_i\left(f^0(R_i,h){u_j^0}'(R_i)-{u_j^0}(R_i)\partial_rf^0(R_i,h)-u_j(R_i)X_j(f^0)(R_i,h)\right)
\d vol_K\\
 &-&\sum_{i=1}^N\int_{h\in K}M_i
\left({u_j^0}(R_i)\partial_rf^\zeta(R_i,h)\zeta^j
+u_jX_j(f^\zeta)(R_i,h)\zeta^j-f^0(R_i,h)u_j'(R_i)\right)\d
vol_K
\,.
\end{eqnarray*}
Before we proceed, we note that while we were dealing with the
circle, terms  such as $u_\theta\partial_\theta f$ disappeared
when we integrated over the circle, as one would expect from
Stokes' theorem. Here, however, we have the terms $u_jX_j(f^0)$
and $u_jX_j(f^\zeta)$. One asks, ``Do these vanish when we integrate over
the group $K$?'' As w shall see, the answer to this question is, ``Yes.''

From standard Riemannian geometry, one has
\begin{prop1}
For any compact Riemannian manifold $N$, and any vector field $X$
and function $\phi$ on $N$ we have
\[
\int_N X(\phi)\d vol=\int_N \phi\,\mathrm{div}\,X\d vol
\,.\]
\end{prop1}
From this formula we see that, in the present situation,
\begin{prop1}
Given any $\kappa\in\mathfrak{k}$ and any function $\phi$ on $K$.
\[
\int_K X_\kappa(\phi) \d vol_K=0
\,.
\]
\end{prop1}
This relation follows because $X_\kappa$ generates volume preserving
(metric preserving!) diffeomorphisms of $K$. 

Hence, upon 
integrating over $K$, the terms $u_jX_i(f^0)$ and $u_jX_j(f^\zeta)$
will vanish because $u_j$ is $K$-invariant. Thus, we have
\begin{eqnarray*}
\int_\Sigma\bm^j([w,\bu_j])&=&\sum_{i=1}^N\int_{h\in
K}P_i\left(f^0(R_i,h){u_j^0}'(R_i)-{u_j^0}(R_i)\partial_rf^0(R_i,h)\right)
\d vol_K\\
 &-&\sum_{i=1}^N\int_{h\in K}M_i
\left({u_j^0}(R_i)\partial_rf^\zeta(R_i,h)\zeta^j-f^0(R_i,h)u_j'(R_i)\right)\d
vol_K
\end{eqnarray*}
This formula implies the EPDiff equations,
\begin{eqnarray*}
0&=&\int_\Sigma\left(\diff{\bm^j}{t}(w)+\bm^j([w,\bu_j])\right) \d
vol_p\\
 &=&\sum_{i=1}^N\int_{h\in K}
 \left(\dot{P}_i f^0(R_i,h)+P_i\dot{R}_i\Diff{R_i}f^0(R_i,h)+\dot{M}_i f^\zeta(R_i,h)\zeta^j+M_i\dot{R}_i\Diff{R_i}
f^\zeta(R_i,h)\zeta^j\right)\d vol_K\\
 &+&\sum_{i=1}^N\int_{h\in
K}P_i\left(f^0(R_i,h){u_j^0}'(R_i)-{u_j^0}(R_i)\partial_rf^0(R_i,h)\right)
\d vol_K\\
 &-&\sum_{i=1}^N\int_{h\in K}M_i
\left({u_j^0}(R_i)\partial_rf^\zeta(R_i,h)\zeta^j-f^0(R_i,h)u_j'(R_i)\right)\d
vol_K
\end{eqnarray*}
Again, we compare coefficients of $f^0,f^\zeta,\Diff{R_i}f^0$ and
$\Diff{R_i}f^\zeta$ to find
\begin{eqnarray*}
0&=&\dot{P}_i +P_i{u_j^0}'(R_i)+M_iu_j'(R_i)\,,\\
0&=&P_i\dot{R}_i
-P_i{u_j^0}(R_i)\,,\\
0&=&\dot{M}_i\zeta^j\,,\\
0&=&M_i\dot{R}_i\zeta^j-M_i{u_j^0}(R_i)\zeta^j
\,.\end{eqnarray*}
Therefore, we have proven the following.

\begin{thm1}
[Hamiltonian diffeon reduction for $K$-invariant warped product spaces]
The warped product diffeon parameters in equation (\ref{wdiff}) satisfy
\begin{eqnarray}
\label{gench1} \dot{P}_i&=& 
-\,\frac{1}{2}\Diff{R_i}\sum_{k=1}^N\left(P_k{u_j^0}(R_k)+M_ku_j(R_k)\right)
\,,\\
\label{gench2}\dot{R}_i &=&{u_j^0}(R_i) 
\,,\\
\label{gench3}\dot{M_i}&=&0
\,.
\end{eqnarray}
As guaranteed by the momentum map property of the diffeon solution
ansatz (\ref{singsoln}) for EPDiff, these are Hamilton's equations with
collectivized Hamiltonian $H_j/(\mathrm{vol}K)$ given by
\begin{eqnarray*}
H_j(\mathbf{P},\mathbf{R};\mathbf{M})
&=&\frac{1}{2}\int_\Sigma\bm^j(\bu_j)\d vol_P\\
                        &=&
\frac{1}{2} \mathrm{vol}\,(K)\sum_{i=1}^N\left(P_i u_j^0(R_i)+M_i
u_j(R_i)\right)\\
&=&\frac{1}{2} \mathrm{vol}\,(K)\sum_{i=1}^N \left(P_i
u_j^0(R_i)+M_i u_j(R_i)\right)\\
&=&\frac{1}{2} \mathrm{vol}\,(K)\sum_{i,k=1}^N \left(P_i
P_kG^0(R_i,R_k)+\frac{M_iM_k}{\psi(R_i)\psi(R_k)}G^K(R_i,R_k)\right).\\
\end{eqnarray*}
Notice that $H_j$ is independent of $j$; so we may write $H^K$ for
$H_j$.
\end{thm1}

 Thus, given any $\zeta\in\mathfrak{k}$ a diffeon exists 
that moves on $\Sigma$ with motion described by Hamilton's
equations with the Hamiltonian given by $H$. The diffeon itself is
``rotating" in the direction determined by $\zeta$ with constant
angular momentum determined by $\mathbf{M}=(M_1,\ldots,M_N)$.
\begin{rmk1}
One may repeat the whole procedure replacing $K$ with a compact
symmetric space $K/H$ for some closed Lie subgroup $H$ of $K$. The
only significant changes would be that the
$\theta^i$ would become local on $K/H$ rather than global.
However, all the propositions will remain true because of the
intimate relationship between symmetric spaces and Lie groups.
Thus, for example, we would expect diffeon behaviour on $\sphere{n}$ to be
similar to the Puckon behaviour on $\sphere{2}$ since
\[
g_{\sphere{n}}=\dr^2+(\cos^2r)\, g_{\sphere{n-1}}.
\]
\end{rmk1}
\subsection{Singular fibres}
We have so far dealt with a free action of a group on a manifold
such that the quotient space is a 1-manifold. Thinking back to the
case of $\Sigma=\sphere{2}$, we see that the action of the circle was not 
completely free, because there are precisely two points (the poles) which
are fixed under the group. Away from these fixed points, the sphere is a
principal $\sphere{1}$ fibration, and the Puckons ``bounce" off
the poles (provided they are not rotating). For the general manifold
$\Sigma$, the situation could become much more complex.\par For example,
in the situation of the previous section, we see that problems arise if
we choose a diffeon which is rotating in the direction
$\zeta\in\mathfrak{k}$ and find that $X_\zeta$ vanishes at a point
$p$. However, if the diffeon is not rotating, and the vector field
$\Diff{r}$ vanishes nowhere, then all the previous theory holds.
The theory also holds for the warped product of the line (or
circle) with the flat metric with an Einstein space, upon using harmonic
coordinate charts (p285 of \cite{P}).

%

\section{Conclusions}
We provided a canonical Hamiltonian framework for exploring the solutions of
the EPDiff equations on surfaces of constant curvature. This framework used
symmetry and the momentum map for singular solutions to reduce the EPDiff
integro-partial differential equation to canonical Hamiltonian ODEs in time. We
specialized to the case of the sphere and provided both numerical integrations
and qualitative analysis of the solutions, which we called ``Puckons."

The main conclusions from our numerical study were:
\begin{itemize}
\item

Momentum plays a key role in the dynamics of Puckons. Radial momentum drives
Puckons to collapse onto one of the poles, and angular momentum
prevents this collapse from occuring. Puckons were found to
exhibit elastic collision behavior (with its associated exchanges of momentum
and angular momentum, but with no excitation of any internal degrees of freedom)
just as occurs in soliton dynamics. 

\item
Puckons without rotation may collapse onto one of the poles. This collapse 
occurs with bounded canonical momentum and the radial slope in velocity appears
to become vertical at the instant of collapse.
\item
For nonzero rotation, Puckon collapse onto one of the poles cannot occur and 
the radial slope in velocity never becomes infinite.
\item
Head-on collisions between two Puckons may be accompanied by an apparently
vertical radial slope in velocity which forms in finite time.
\end{itemize}
The main theoretical questions that remain are:
\begin{itemize}
\item
Numerical simulations show that near vertical or vertical slope occurs at
head-on collision between two Puckons of nearly equal height. A rigorous proof
of this fact is still missing.
\item
It remains to discover whether a choice of Green's function exists
for which the reduced motion is integrable on our $2N$ dimensional
Hamiltonian manifold of concentric rotating Puckons for $N>1$.
\item
It also remains to determine the number and speeds of the rotating Puckons that
emerge from a given initial condition.
\end{itemize}
All of these challenging theoretical problems are beyond the scope of the
present paper and we will leave them as potential subjects for future work.

We applied these ideas to hyperbolic spaces, as well. This led to rather
simple reduced dynamics with only a limited number of possible collisions.
We suggested a new departure for hyperbolic space, based on Teichm\"uller
theory, which we shall investigate elsewhere.

Finally, we answered an outstanding question by generalising the momentum map
to the case of diffeons with $(n-1)-$dimensional internal degrees of freedom by
using the theory of warped product spaces. 

In summary, we identified and analysed cases where imposing an additional
translation symmetry on the solution reduced the canonical Hamiltonian
dynamics of the singular solutions of EPDiff on Einstein surfaces from
(integral) partial differential equations to Hamilton's canonical ordinary
differential equations, in time. We extended our methods for surfaces to
``mostly symmetric'' manifolds in higher dimensions by using warped products.

\subsection*{Acknowledgements}
DDH is grateful for partial support by US DOE,
under contract W-7405-ENG-36 for Los Alamos National Laboratory, and Office
of Science ASCAR/AMS/MICS. The research of JM was partially supported by an
EPSRC postdoctoral fellowship at Imperial College London.
SNS is supported by a US Department of Energy Computational
Science Graduate Fellowship under grant number DE-FG02-97ER25308.
 The authors wish to thank John Gibbon, Peter Lynch, and Richard Thomas
  for their thoughts and advice.

\rem{\bibliography{Bibtex}
 \bibliographystyle{ACM}}

\end{document}